\documentclass[reprint,showpacs,preprintnumbers,amsmath,amssymb,aps,prd,floatfix,longbibliography,superscriptaddress]{revtex4-2}

\usepackage{bbold}
\usepackage{hyperref}

\usepackage{graphicx}

\usepackage{color}

\begin{document}

\title{Measuring gravitational time dilation with delocalized quantum superpositions}

\author{Albert Roura}
%\email{}
\affiliation{Institute of Quantum Technologies, German Aerospace Center (DLR),
S\"oflinger Stra\ss e 100, 89077 Ulm, Germany}
\author{Christian Schubert}
\affiliation{Institut f\"ur Quantenoptik, Leibniz Universit\"at Hannover,
Welfengarten 1, D-30167 Hannover, Germany.} 
\affiliation{Institute for Satellite Geodesy and Inertial Sensing, German Aerospace Center (DLR),\\
c/o Leibniz Universit\"at Hannover, DLR-SI, Callinstra\ss e 36, 30167 Hannover, Germany}
\author{Dennis Schlippert}
\affiliation{Institut f\"ur Quantenoptik, Leibniz Universit\"at Hannover, Welfengarten 1, D-30167 Hannover, Germany.} 
\author{Ernst~M.~Rasel}
\affiliation{Institut f\"ur Quantenoptik, Leibniz Universit\"at Hannover, Welfengarten 1, D-30167 Hannover, Germany.} 

\date{\today}

\begin{abstract}
Atomic clocks can measure the gravitational redshift predicted by general relativity
with great accuracy and for height differences as little as $1\, \text{cm}$. All existing experiments, however, involve the comparison of two independent clocks at different locations rather than a single clock in a delocalized quantum superposition.
Here we present an
interferometry scheme employing group-II-type atoms, such as Sr or Yb, capable of measuring the gravitational time dilation in a coherent superposition of atomic wave packets at two different heights.
In contrast to other recent proposals, there is no need for pulses that can efficiently diffract both internal states.
Instead, the scheme relies on very simple atom optics for which high-diffraction efficiencies can be achieved with rather mild requirements on laser power.
Furthermore, the effects of vibration noise are subtracted by employing a simultaneous Rb interferometer that acts as an inertial reference.
Remarkably, the recently commissioned VLBAI facility in Hannover, a 10-meter atomic fountain that can simultaneously operate Yb and Rb atoms and enables up to $2.8\, \text{s}$ of free evolution time, meets all the requirements for a successful experimental implementation.
\end{abstract}


\maketitle

%%%%%
%\section{Introduction and quantum-clock model}
%%%%%

\subsection*{\large Introduction}

Light-pulse atom interferometers \cite{kasevich91,kleinert15} can create quantum superpositions of atomic wave packets with spatial separations reaching the half-meter scale \cite{kovachy15b}, and have shown a great potential as inertial sensors \cite{peters99,savoie18} for both practical applications
and highly sensitive measurements in fundamental physics \cite{bongs19}. The latter include the accurate determination of fundamental constants \cite{rosi14,bouchendira11,parker18} as well as high-precision tests of QED \cite{bouchendira11,parker18}, the universality of free fall (UFF) 
\cite{schlippert14,zhou15,rosi17b,asenbaum20} and certain dark-energy models \cite{hamilton15b,jaffe17,sabulsky19}.

On the other hand, the remarkable accuracy achieved by atomic clocks \cite{bloom14,campbell17,ludlow15}, which has been exploited in searches of ultralight dark-matter candidates \cite{hees16} and of tiny violations of Lorentz invariance \cite{matveev13,pruttivarasin15}, enables measurements of the gravitational redshift that have confirmed the agreement with Einstein's predictions to one part in $10^5$ \cite{delva18,hermann18} and for height differences as small as $1\, \text{cm}$ \cite{chou10,takamoto20}.
So far these experiments have always relied on the comparison of several independent clocks. Nevertheless, in order to investigate general relativistic effects in a truly quantum regime, it would be of great interest to measure the effects of gravitational time dilation for a single clock in a quantum superposition of 
wave packets peaked at different heights.

Contrary to initial claims \cite{mueller10}, commonly employed 
atom interferometers cannot be exploited to measure the gravitational redshift \cite{wolf11a,schleich13a}. In fact, even quantum-clock interferometry experiments \cite{sinha11} where atoms are prepared in a superposition of two internal states
and then sent through a light-pulse atom interferometer
are insensitive to gravitational time-dilation effects in a uniform gravitational field \cite{roura20a,loriani19}.
As shown in Ref.~\cite{roura20a}, on the other hand, this lack of sensitivity 
can be overcome by initializing the quantum clock when the spatially separated superposition of atomic wave packets has already been generated.
However, that scheme (and related ones \cite{ufrecht20}) involve laser pulses capable of 
efficiently diffracting both internal states. And such pulses entail very demanding requirements on laser power or rather complex set-ups for implementing new diffraction techniques 
that have not been demonstrated yet and will need 
years of further development.
Moreover, some of these interferometer configurations \cite{ufrecht20} 
are also rather sensitive to vibration noise, so that long integration times are necessary in order to average its effects out.

In contrast, we will present here an alternative interferometry scheme that is directly sensitive to the gravitational redshift and relies on simple atom optics with very mild requirements on laser power.
Furthermore, the effects of vibration noise can be subtracted by employing a simultaneous
interferometer for a second atomic species 
that acts as an inertial reference.
The emphasis of this proposal is on the feasibility and simplicity of its practical implementation. Indeed, we expect the requirements to be met by the recently commissioned VLBAI facility in Hannover, a 10-meter atomic fountain that will enable up to $2.8\, \text{s}$ of free evolution time and simultaneous operation of Yb and Rb atoms \cite{hartwig15,schlippert20}.

\subsection*{\large Quantum-clock model}

We will consider the group-II-type atoms typically employed in optical atomic clocks, and for our purposes it will be sufficient to focus just on the two clock states.
Therefore, as far as their internal state is concerned, they can be suitably modeled as a two-level system involving the ground state $| \text{g} \rangle$ and an excited state $| \text{e} \rangle$ with an additional energy $\Delta E$. Due to this energy difference, if one prepares at some time $\tau_\text{i}$ a linear superposition
$\big( | \mathrm{g} \rangle + | \mathrm{e} \rangle \big) / \sqrt{2}$
of the two energy eigenstates, the relative phase between them will grow with time and encode the elapsed proper time $\Delta \tau$:
\begin{equation}
\big| \Phi (\tau) \big\rangle \propto \frac{1}{\sqrt{2}}
\left( | \mathrm{g} \rangle + e^{- i \Delta E \, \Delta\tau / \hbar} | \mathrm{e} \rangle \right)
\label{eq:evolution1} ,
\end{equation}
where $\Delta \tau = \tau - \tau_\text{i}$, and the left- and right-hand sides are equal up to a global phase factor.

In addition, however, one needs to consider %(the quantum dynamics of)
the atom's center-of-mass (COM) degree of freedom.
The quantum state of the atom is therefore an element of a Hilbert space given by the tensor product of the internal space and the Hilbert space associated with the COM:
\begin{equation}
\big| \Psi \big\rangle = \big| \psi^{(1)} \big\rangle \otimes | \mathrm{g} \rangle
+ \big| \psi^{(2)} \big\rangle \otimes | \mathrm{e} \rangle
\label{eq:state1} .
\end{equation}
In the absence of laser pulses its time evolution is generated by the following
Hamiltonian operator:
\begin{equation}
\hat{H} = \hat{H}_1 \otimes | \mathrm{g} \rangle \langle \mathrm{g} | + \hat{H}_2 \otimes | \mathrm{e} \rangle \langle \mathrm{e} |
\label{eq:hamiltonian1} ,
\end{equation}
where  $\hat{H}_1$ and $\hat{H}_2$ are the Hamiltonians that govern the dynamics %free evolution
of a relativistic particle with rest mass $m_1$ and $m_2 = m_1 + \Delta m$ respectively. The additional rest mass $\Delta m = \Delta E / c^2$ accounts for the energy difference
between the two internal states.

As shown in Ref.~\cite{roura20a}, and generalizing previous results in the non-relativistic case \cite{borde92,antoine03b,hogan08,roura14}, the propagation of an atomic wave packet in curved spacetime can be described in terms of a worldline $X^\mu (\tau)$ corresponding to its \emph{central trajectory} and the evolution of a \emph{centered wave packet}
accounting for the wave packet's expansion and shape evolution.
It is particularly useful to consider a comoving frame where $X^\mu(\tau_\text{c}) = \big(c\, \tau_\text{c}\, , \mathbf{0} \big)$ and the comoving time coordinate $\tau_\mathrm{c}$ coincides with the proper time along the worldline. In this frame the Hamiltonians $\hat{H}_1$ and $\hat{H}_2$ take a simple form involving two contributions: a $c$-number corresponding to the rest mass energy $m_n c^2$, and a second contribution $\hat{H}_\mathrm{c}^{(n)}$ that governs the dynamics of the centered wave packet $\big| \psi_\text{c}^{(n)} (\tau_\text{c}) \big\rangle$ and reduces to the Hamiltonian of a non-relativistic particle, provided that
the wave-packet size and its velocity spread are much smaller than the spacetime curvature radius and the speed of light, respectively. 
The evolution of the COM state associated with each internal state can then be written as
\begin{equation}
\big| \psi^{(n)} (\tau_\text{c}) \big\rangle = e^{i\hspace{0.2ex} \mathcal{S}_n / \hbar}\,
\big| \psi_\text{c}^{(n)} (\tau_\mathrm{c}) \big\rangle
\label{eq:state_evol1} ,
\end{equation}
where the index $n = 1,2$ 
labels the internal state and $\mathcal{S}_n$ is the \emph{propagation phase}, which is given for freely falling atoms by the rest mass energy times the proper time along the central trajectory:
\begin{equation}
\mathcal{S}_n = - m_n c^2 \int^{\tau}_{\tau_0} d\tau'
\label{eq:prop_phase1} .
\end{equation}
Further details can be found in Ref.~\cite{roura20a}, where a relativistic description of atom interferometry in curved spacetime applicable to a wide range of situations, including also the effects of any external forces and guiding potentials, has been developed. 

In the interferometers that we will be considering the evolution of the centered wave packets is
the same along the two arms, and it will be sufficient to focus on the central spacetime trajectories, which can be conveniently displayed in simple spacetime diagrams, and the proper time along those. Moreover, since they are invariant geometric quantities,
when calculating these proper times, one can choose any particularly convenient 
coordinate system.
In this way, for weak gravitational fields and non-relativistic velocities the calculation reduces to evaluating the classical action:
\begin{equation}
\mathcal{S}_n \approx \int^{t}_{t_0} dt' \left( - m_n\, c^2 + \frac{1}{2} m_n \dot{\mathbf{X}}^2
- m_n\,  U (t',\mathbf{X}) \right)
\label{eq:prop_phase2} ,
\end{equation}
where the parametrization $X^\mu(t') = \big(c\, t', \mathbf{X}(t') \big)$ in terms of the usual coordinates in a post-Newtonian expansion \cite{misner73} has been employed for the central trajectory,
the overdot denotes a derivative with respect to $t'$ and $U (t',\mathbf{x})$ is the gravitational potential.
In particular, for a uniform gravitational field the 
potential is simply given by $U (t',\mathbf{x}) = U_0 - \mathbf{g} \cdot (\mathbf{x} - \mathbf{x}_0)$.

%%%%
%\section{Quantum-clock interferometry}
%%%%
\subsection*{\large Quantum-clock interferometry}

A natural way of observing 
time-dilation effects in delocalized quantum superpositions is by performing a quantum-clock interferometry experiment \cite{sinha11} with the same kind of atoms employed in optical atomic clocks, such as Sr or Yb. In this case, one prepares an equal-amplitude superposition of the two internal clock states which is then used as the initial state of a light-pulse atom interferometer, where the atomic wave packet is split, redirected and finally recombined by a series of laser pulses acting as diffraction gratings. As emphasized 
in Ref.~\cite{zych11}, any differences in the time dilation along the two arms lead to a contrast reduction of the interferometric signal. However, this effect is far too small to be observable within the parameter regimes accessible to current experiments \cite{roura20a}. Furthermore, this kind of interferometers are insensitive to gravitational time-dilation in a uniform field.
This lack of sensitivity 
can be easily understood by considering a freely falling frame \cite{roura20a}, where the central trajectories correspond to straight lines independent of the gravitational acceleration $\mathbf{g}$, and has also been explicitly shown in a non-relativistic calculation \cite{loriani19}.

\begin{figure}[h]
\begin{center}
\includegraphics[width=8.5cm]{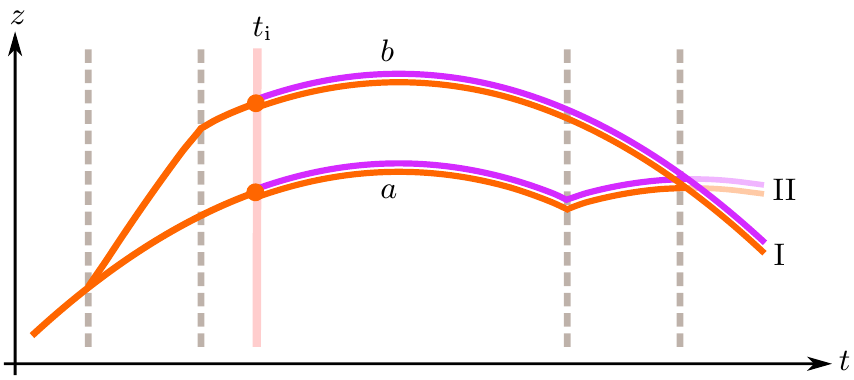}\\
\vspace{6.0ex}
\includegraphics[width=8.5cm]{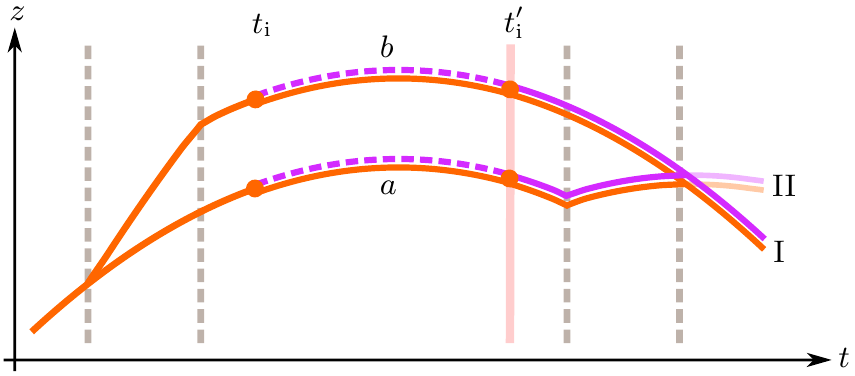}
%\vspace{-0.5ex}
\end{center}
\caption{Central trajectories for a reversed Ramsey-Bord\'e interferometer \cite{borde89},
which involves two pairs of laser pulses acting as diffraction gratings (grey dashed lines),
in a uniform gravitational field. A differential phase-shift measurement of the ground (orange) and excited (purple) states is performed for various initialization times ($t_\text{i}$~and~$t'_\text{i}$). Comparison of the outcomes for the two different initialization times is directly related to the proper-time difference between the dashed segments in the two arms ($a$ and $b$), which is a consequence of gravitational time dilation.}
\label{fig:doubly_differential}
\end{figure}

As recently proposed \cite{roura20a}, these difficulties can be circumvented by initializing the clock (i.e.\ generating the superposition of internal states) after the superposition of spatially separated wave packets has already been created and then performing a state-selective measurement of the exit ports in order to determine the interferometer phase shift for each of the two internal states.
The differential phase shift between the two states contains in that case very valuable information.
In fact, a doubly differential measurement comparing the outcomes of the differential measurements for two different initialization times $t_\mathrm{i}$ and $t_\mathrm{i}'$, as illustrated in Fig.~\ref{fig:doubly_differential}, is directly related to the gravitational redshift between the two arms.
Indeed, the difference between the two differential measurements corresponds to the additional time spent in the excited state for the earlier initialization (dashed segments) as well as the different gravitational time dilation for the two arms during that period due to the height difference.

An important aspect of the scheme of Fig.~\ref{fig:doubly_differential} is that the phase shift for both internal states is simultaneously measured in a single shot through state-selective detection. This is because the differential phase-shift measurement benefits from common-mode rejection of unwanted effects acting commonly on both internal states, and the simultaneous measurement guarantees that such cancelation also holds for effects that are not stable from shot to shot such as vibration noise of the retroreflection mirror, which is otherwise the typical dominant noise source for long interferometer times.
However, the main challenge of such a scheme is that the diffraction pulses applied after the initialization pulse should be capable of efficiently diffracting atoms in either of the two internal states and should actually have comparable Rabi frequencies in both cases.

A natural option for the simultaneous diffraction of both internal states is Bragg diffraction \cite{kozuma99} at the magic wavelength \cite{ye08}, which guarantees that the optical potentials, and hence the Rabi frequencies, are indeed the same for $| \text{g} \rangle$ and $| \text{e} \rangle$. This wavelength is, however, far detuned from any transition and requires rather large laser intensities in order to achieve Rabi frequencies that are not too low. Lower Rabi frequencies require longer pulses and lead to reduced diffraction efficiencies due to higher velocity selectivity \cite{szigeti12}, which becomes a serious limiting factor even for atomic clouds with narrow momentum distributions.
Furthermore, the spatial extent of atomic clouds freely evolving for several seconds also constrains the minimum beam size \cite{abend20}, which altogether places very demanding requirements on laser power.
Indeed, this is clearly illustrated by the following quantitative example for Yb atoms: $5\, \text{W}$ of laser power and a 1-cm beam waist lead to a Rabi frequency $\Omega \approx 2\pi \times 11\, \text{Hz}$,
and even for a narrow momentum distribution with $T_\text{eff} = 1\, \text{nK}$ 
such a Rabi frequency would imply a diffraction efficiency for a single $\pi/2$ pulse of less than 3\% compared to an ideal pulse.

An alternative diffraction mechanism proposed in Ref.~\cite{roura20a} involves a combination of simultaneous pairs of pulses driving single-photon transitions between the two clock states. The application of these single-photon transitions to atom interferometry has already been demonstrated for $^{88}\mathrm{Sr}$ atoms \cite{hu17,hu20}, but a large magnetic field was necessary to turn the otherwise forbidden transition for bosonic isotopes into a weakly allowed one. Since the use of such magnetic fields does not seem viable for high-precision measurements, fermionic isotopes, which are harder to cool down to the required ultra-low temperatures, will need to be employed instead. Substantial efforts in this direction are expected in the near future because several large-scale projects \cite{magis,badurina20} will rely on atom interferometers based on such transitions. However, a number of years of further development will still be necessary to reach the required maturity level. Furthermore, using this kind of pulses involves more sophisticated set-ups and frequency stabilization methods. 
Instead, simpler diffraction techniques available to any laboratory working on light-pulse atom interferometry, such as standard Bragg diffraction, would be desirable.

%%%%%
%\section{Alternative scheme}
%%%%%
\subsection*{\large Alternative interferometry scheme}

\begin{figure}[h]
\begin{center}
\includegraphics[width=8.5cm]{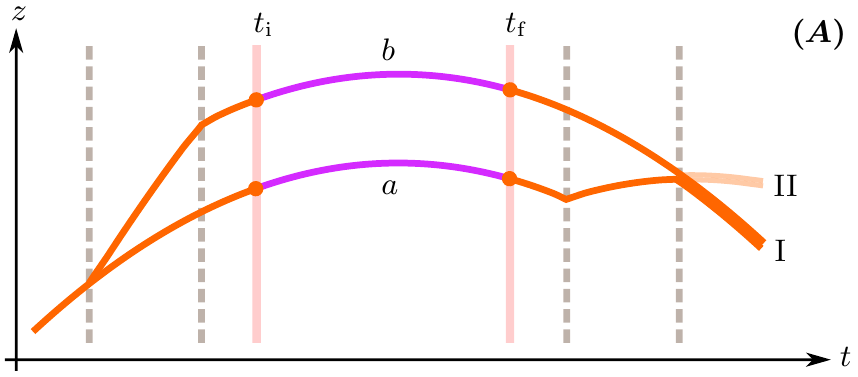}\\
\vspace{6.0ex}
\includegraphics[width=8.5cm]{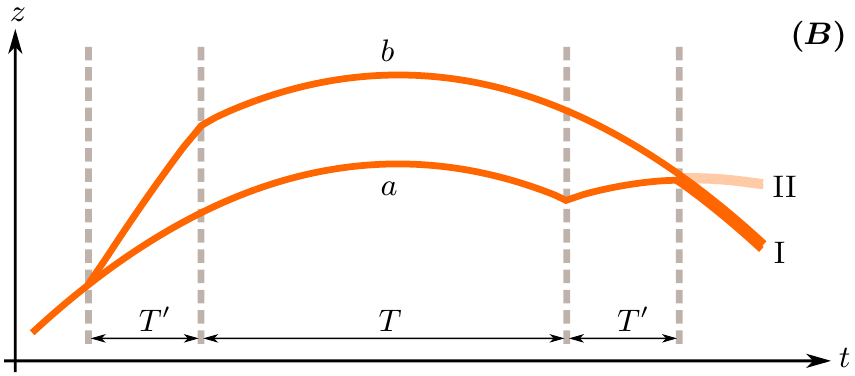}
%\vspace{-0.5ex}
\end{center}
\caption{Central trajectories for a reversed Ramsey-Bord\'e interferometer in the laboratory frame. In this frame the inversion pulses applied at times $t_\text{i}$ and $t_\text{f}$ act simultaneously on both arms ($A$). Nevertheless, the proper time spent in the excited state (purple) is slightly different for the two arms due to gravitational time dilation. 
By repeating the measurement without inversion pulses ($B$) and subtracting the phase shifts obtained in the two cases, one gets the same result 
as in the quantum-clock interferometry scheme of Fig.~\ref{fig:doubly_differential}.}
\label{fig:inversions}
\end{figure}

In order to address these challenges
and look for alternatives involving much simpler atom optics, let us consider the possibility of measuring the phase shift accumulated by the two internal states in separate shots rather than simultaneously. This can be accomplished in two shots as shown in Fig.~\ref{fig:inversions}.
First, in one shot~($A$)
the initialization pulse (a $\pi/2$ pulse) at time $t_\mathrm{i}$ is replaced with an inversion pulse (a $\pi$ pulse) that swaps the internal state from $| \text{g} \rangle$ to $| \text{e} \rangle$ instead of generating an equal-amplitude superposition. Furthermore, at some later time $t_\mathrm{f}$ in the same shot one
applies a second inversion pulse that swaps the internal state back to $| \text{g} \rangle$. Next, one repeats the measurement in a subsequent shot~($B$)
with no inversion pulses but the same laser-pulse sequence otherwise.
The differential phase shift between the two shots is then given by the proper time spent by the atoms in the excited state and how it differs for the two arms ($a$ and $b$) due to the gravitational redshift:
\begin{align}
\delta \phi_A - \delta \phi_B &= 
- \Delta m\, c^2  \, (\Delta\tau_b - \Delta\tau_a) / \hbar
\nonumber \\
%= - \frac{\Delta E}{2 \hbar} \, (\Delta\tau_b - \Delta\tau_a)
&= - \Delta m \, g \, \Delta z \, (t_\text{f} - t_\text{i}) / \hbar
\label{eq:inversions}\, ,
\end{align}
where $\Delta z$ is the vertical separation between the two arms,
and $g$ is the gravitational acceleration along the $z$ direction, which coincides with the direction of the laser beams.
Interestingly, in contrast with the scheme depicted in Fig.~\ref{fig:doubly_differential}, this is achieved with a single differential measurement.

\begin{figure}[t]
\begin{center}
\vspace{2.0ex} %%%
\includegraphics[width=8.5cm]{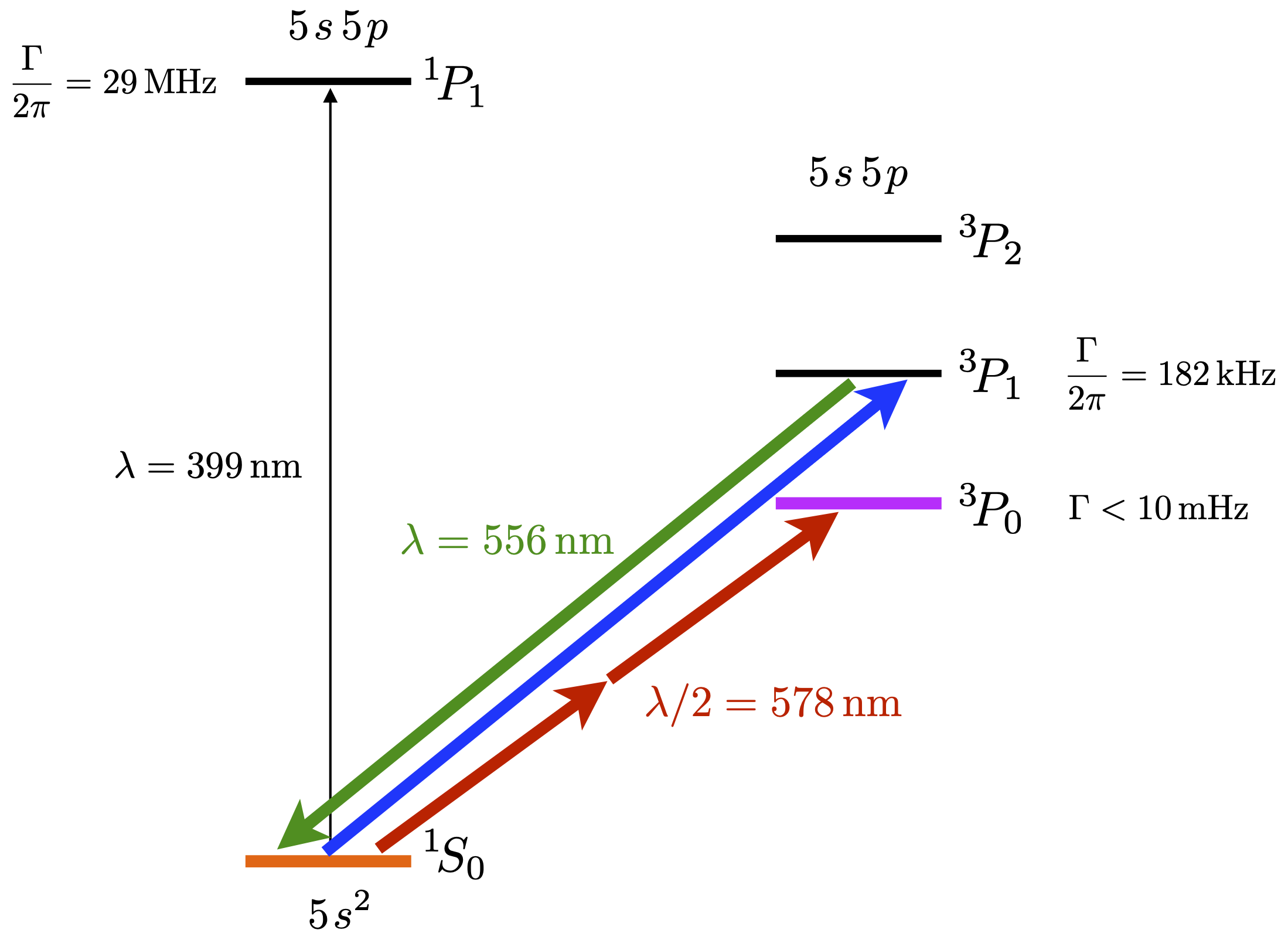}
\vspace{-1.5ex}
\end{center}
\caption{Relevant energy levels and transitions for 
Yb atoms.
The two-photon E1--M1 transition (red arrows) employed for the inversion pulses can be interpreted as the absorption of two counterpropagating equal-frequency photons.
On the other hand, the Bragg transition (blue and green arrows) on which the diffraction pulses are based involves the absorption and stimulated emission of two counterpropagating photons with a slight  frequency difference $\Delta\omega / 2\pi$ that accounts for the change of the atom's kinetic energy.
The two clock states are respectively indicated with orange and purple color.}
\label{fig:transitions}
\end{figure}

\begin{figure}[h]
\begin{center}
\includegraphics[width=4.5cm]{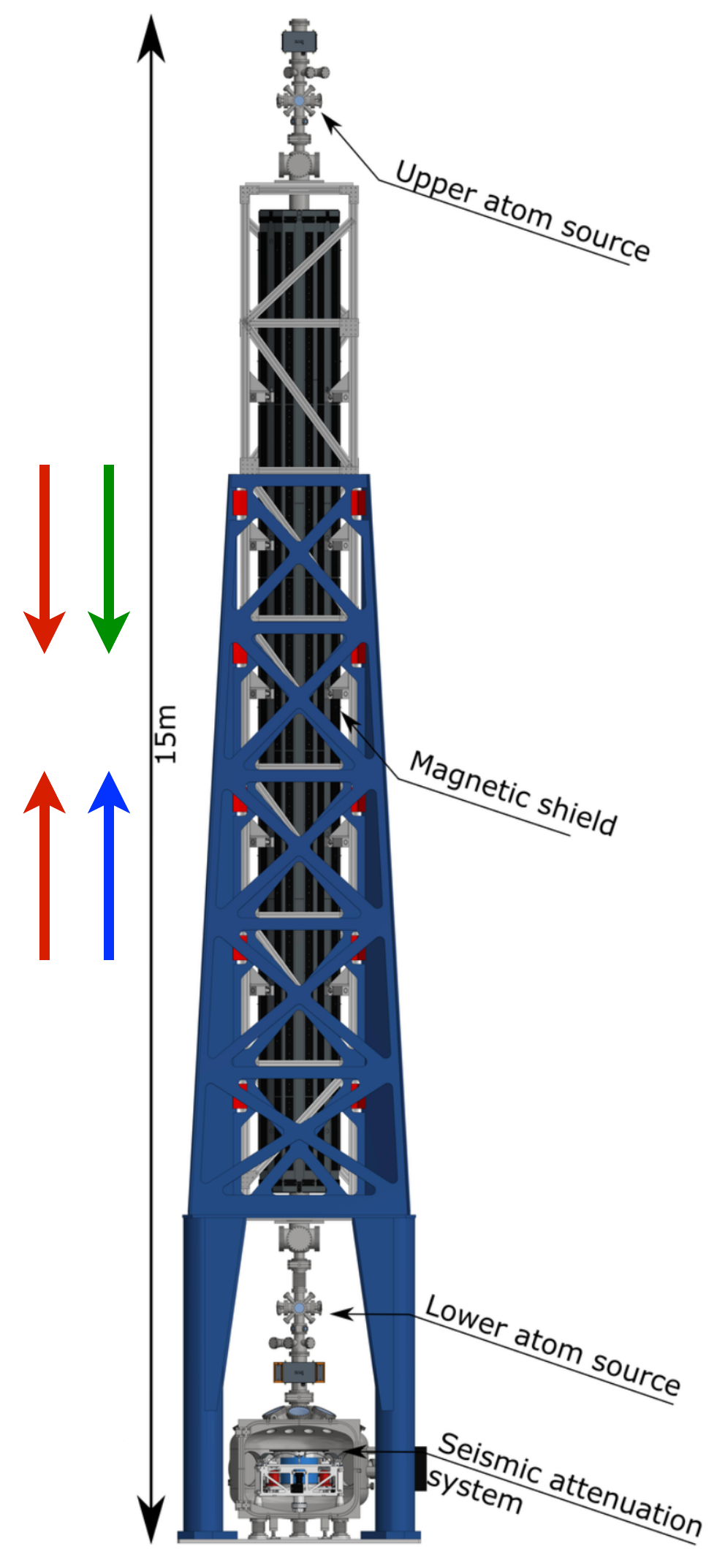}
\vspace{-0.5ex}
\end{center}
\caption{Schematic diagram of the VLBAI facility, a 10-meter atomic fountain capable of simultaneously operating Rb and Yb atom interferometers. The directions of the laser beams employed for the inversion  (red) and Bragg diffraction (green and blue) pulses are also displayed. The upward propagating beams have been retroreflected by the vibrationally isolated mirror at the bottom.}
\label{fig:VLBAI}
\end{figure}

More importantly, the application of the second inversion pulse implies that the second pair of diffraction pulses also act on atoms in the ground state, which means that all the diffraction pulses in both shots act on the same internal state, and the challenge associated with the diffraction of different internal states is entirely overcome.
In fact, an efficient diffraction mechanism for Sr and Yb atoms in the ground state has already been demonstrated in Refs.~\cite{del_aguila18,plotkin-swing18} and is readily available. It employs Bragg diffraction based on the $^1S_0 - {}^3P_1$ intercombination transition, a two-photon process schematically indicated with blue and green arrows in Figs.~\ref{fig:transitions} and \ref{fig:VLBAI}.
Alternatively, Bragg diffraction based on the $^1S_0 - {}^1P_1$ transition is also possible \cite{mazzoni15}.
Furthermore, large momentum transfer (LMT), which allows reaching higher $\Delta z$ with shorter times $T'$ and leaving more time available for $(t_\text{f} - t_\text{i})$, was demonstrated in both cases too.
Finally, as done in Refs.~\cite{del_aguila18,plotkin-swing18,mazzoni15}, these diffraction pulses can also be used with bosonic isotopes, which have a simpler spectroscopic structure that makes them simpler to cool and can even reach Bose-Einstein condensation \cite{roy16,plotkin-swing18}.

Besides the diffraction pulses, the inversion pulses also play a central role. They are based on the two-photon E1--M1 transition between the two clock states investigated in Ref.~\cite{alden14}. It is indicated with red arrows in Figs.~\ref{fig:transitions} and \ref{fig:VLBAI}, and employs equal-frequency counterpropagating photons whose frequencies equal half the frequency of the clock transition. In contrast to the single-photon case, they can drive the transition between the clock states for bosonic atoms without the need for strong magnetic fields to be applied, and have three important and closely related properties.
First, in the laboratory frame the transition implies no momentum transfer to the atomic wave packet. Second, this is a Doppler-free transition, i.e. corrections to the resonance condition due to the initial velocity of the atoms cancel out at linear order and only much smaller terms quadratic in the velocity contribute.
Finally, the spatial dependence of the effective phase associated with the two-photon process cancels out in the laboratory frame and the hypersurfaces of constant phase correspond to simultaneity hypersurfaces in this frame, which is crucial to guarantee the proper synchronization of the inversion pulses acting on the two interferometer arms. In fact, due to the relativity of simultaneity for spatially separated events, these pulses do not act simultaneously on both arms in the freely falling frame \cite{roura20a}, and this is actually how the insensitivity to gravitational redshift for a uniform field is circumvented in this case.

%\begin{figure}[ht]
%\begin{center}
%\includegraphics[width=8.5cm]{figures/figure3.pdf}
%\vspace{-1.5ex}
%\end{center}
%\caption{Relevant energy levels and transitions for %\comment{bosonic}
%Yb atoms.
%The two-photon E1--M1 transition \comment{(red arrows)} employed for the inversion pulses can be interpreted as the absorption of two counterpropagating equal-frequency photons.
%On the other hand, the Bragg transition \comment{(blue and green arrows)} on which the diffraction pulses are based involves the absorption and stimulated emission of two counterpropagating photons with a slight  frequency difference \comment{$\Delta\omega / 2\pi$} that accounts for the change \comment{of} the atom's kinetic energy.}
%\label{fig:transitions}
%\end{figure}

As shown by Eq.~\eqref{eq:inversions}, the new interferometry scheme displayed 
in Fig.~\ref{fig:inversions} can successfully measure the difference of gravitational time dilation between the two arms during the free evolution 
between the two inversion pulses.
In order to confirm that these measurements can indeed be interpreted as tests of the universality of gravitational redshift (UGR) with macroscopically delocalized quantum superpositions, we will follow the approach of Ref.~\cite{roura20a} and consider a dilaton model \cite{damour12,damour10a} as a consistent framework for parametrizing violations of the equivalence principle.
For weak gravitational fields the effect of the dilaton field amounts to
replacing the potential term in Eq.~\eqref{eq:prop_phase2} with $m_n (1 + \beta_n)\, U (t',\mathbf{X})$, where the parameters $\beta_n$ encode deviations from a metric theory of gravity and are directly related to the E\"otv\"os parameter $\eta_\text{e-g}$ characterizing the violations of UFF for the two internal states: $\eta_\text{e-g} \approx (\beta_2 - \beta_1)$. With this new propagation phase one can rederive the result for the differential phase shift in Eq.~\eqref{eq:inversions}
and find that the proper-time difference $\Delta\tau_b - \Delta\tau_a$ is replaced by 
\begin{equation}
\Delta \bar{\tau}_b - \Delta \bar{\tau}_a
\approx \big(1 + \alpha_\text{e-g} \big)\, \big( g\, \Delta z / c^2 \big)\, (t_\text{f} - t_\text{i})
\label{eq:inversions2} ,
\end{equation}
where $\alpha_\text{e-g}$ parametrizes the violations of UGR and is given by
\begin{equation}
\alpha_\text{e-g} = \frac{m_1}{\Delta m} \big( \beta_2 - \beta_1 \big)
\label{eq:alpha} ,
\end{equation}
which reveals the close connection between UGR and UFF \footnote{This relation has previously been found for the comparison of independent clocks by employing energy conservation arguments (and Lorentz invariance) \cite{nordtvedt75,wolf16}.}.
The result in Eqs.~\eqref{eq:inversions2} and \eqref{eq:alpha} coincides with what is obtained for the comparison of two independent clocks with a height difference $\Delta z$. In this case, however, a single clock is in a quantum superposition of two spatially separated wave packets.

\subsubsection*{Suppression of vibration noise}

The alternative interferometry scheme of Fig.~\ref{fig:inversions}
can be regarded as the result of splitting up
the quantum-clock interferometry measurement depicted in Fig.~\ref{fig:doubly_differential},
so that the phase shifts for the two internal states are separately determined in two different shots. While having two separate shots enables the use of a second inversion pulse that eliminates the need for diffraction pulses that act efficiently on both internal states, which is the key advantage of the new scheme, it also implies losing the common-mode rejection of unwanted effects that change from shot to shot
in the differential phase shift measurement.
This is particularly relevant for vibration noise and for small changes of the gravitational field from shot to shot.
Indeed, besides the propagation phase in Eq.~\eqref{eq:prop_phase1} the atomic wave packets acquire an additional phase $\pm \big(\mathbf{k}_\mathrm{eff} \cdot \mathbf{X}(t_j) + \varphi_j \big)$ when diffracted by a laser pulse, and this phase depends on their central position $\mathbf{X}(t_j)$ at that time as well as the laser phase $\varphi_j$, which in turn depends on the position of the retroreflection mirror.
After adding up the phases for all the laser pulses, the resulting phase-shift contribution for the reversed Ramsey-Bord\'e interferometers of Figs.~\ref{fig:inversions}A and \ref{fig:inversions}B is given by:
\begin{align}
\delta \phi_\mathrm{laser} &= \delta \bar{\varphi}
+ \mathbf{k}_\mathrm{eff} \cdot \mathbf{g}\, (1 + \beta_1)\, T' (T + T') \nonumber \\
&\quad + \mathbf{k}_\mathrm{eff} \cdot \Delta \mathbf{g}\, T' (T + T')
- \sum_j \delta \mathbf{k}_\text{eff}^{(j)} \cdot \mathbf{X}_\text{mirror} (t_j)
\label{eq:laser_phase_shift} ,
\end{align}
where $\Delta \mathbf{g}$ accounts for small variations of the gravitational field from shot to shot \footnote{For simplicity we have assumed that $\Delta \mathbf{g}$ was time independent during each shot. Otherwise its contribution to Eq.~\eqref{eq:laser_phase_shift} would be replaced by a linear combination of double time integrals, but all the relevant conclusions would remain unchanged.}, $\delta \bar{\varphi}$ includes the contributions of the
injected laser phases (before retroreflection),
$\mathbf{X}_\text{mirror} (t_j)$ is the position of the retroreflection mirror when the $j$th pulse is applied, and $\delta \mathbf{k}_\text{eff}^{(j)}$ is the difference of momentum transfer between both arms for that pulse.
Note that we have also included the parameter $\beta_1$ introduced above, which parametrizes violations of UFF, and have neglected terms of order $\beta_1\, \Delta \mathbf{g}$.

In the absence of laser phase noise, the contribution of the injected laser phases is given by $\delta \bar{\varphi} = - \Delta \dot{\omega}\, T' (T + T')$, where $\Delta \dot{\omega}$ denotes the chirp rate of the frequency difference $\Delta \omega$ between the two counterpropagating beams, which is employed to compensate the time-dependent Doppler effect for freely falling atoms and guarantee that the various Bragg pulses stay on resonance
(for $\lambda = 556\, \text{nm}$ it amounts to $\Delta \dot{\omega} / 2 \pi \approx 35\, \text{MHz/s}$).
In gravimetry measurements this contribution is actually exploited to determine the gravitational acceleration $g$ by scanning the value of the chirp rate in different shots \cite{peters01}.
Furthermore, matching the two phase-shift contributions in this way, which corresponds to taking $\Delta \dot{\omega} = \mathbf{k}_\mathrm{eff} \cdot \mathbf{g}$, relaxes the accuracy requirement for the pulse timings $T$ and $T'$.

The second term on the right-hand side of Eq.~\eqref{eq:laser_phase_shift}
and the stable part of $\delta \bar{\varphi}$
remain the same for different shots and cancel out in the differential phase shift $\delta \phi_A - \delta \phi_B$. However, this is not the case for the last two terms, which capture the effects of small changes of the gravitational field from shot to shot
and the effects of mirror vibrations \footnote{Note that any terms depending on the initial position and velocity of the mirror cancel out and one is just left with those depending on its
acceleration.}.
Fortunately, this drawback can be successfully mitigated by simultaneously operating an atom interferometer for a second species that serves as an inertial reference. As a concrete example, throughout the rest of the paper we will focus on the particular case of Rb atoms, which will be available for simultaneous operation at VLBAI \cite{schlippert20} and for which there is substantial heritage as far as their use in 10-m atomic fountains is concerned \cite{kovachy15b,kovachy15a,asenbaum20}.

Indeed, by employing also a reversed Ramsey-Bord\'e configuration with the same timings between the laser pulses \footnote{If necessary, one could slightly shift all the pulse times of the Rb interferometer  by the same small amount because high-frequency vibrations are typically much better suppressed by the mirror's vibration isolation system.}, its phase shift $\delta \phi_\mathrm{Rb}$ will be dominated by a contribution analogous to that in Eq.~\eqref{eq:laser_phase_shift} but with $\mathbf{k}_\mathrm{eff}^\mathrm{Rb}$ and $\beta_\mathrm{Rb}$ instead of $\mathbf{k}_\mathrm{eff}$ and $\beta_1$. After rescaling by $\big(k_\mathrm{eff} / k_\mathrm{eff}^\mathrm{Rb} \big)$, the contributions of $\Delta \mathbf{g}$ and the mirror vibrations cancel out in the differential phase shift $\big( \delta \phi_A - \big(k_\mathrm{eff} / k_\mathrm{eff}^\mathrm{Rb} \big) \delta \phi_\mathrm{Rb} \big)$. An analogous cancelation holds for shot B, and the gravitational time dilation can thus be directly obtained from the following doubly differential phase shift:
\begin{widetext}
\begin{equation}
\left( \delta \phi_A - \left(\frac{k_\mathrm{eff}}{k_\mathrm{eff}^\mathrm{Rb}}\right) \delta \phi_\mathrm{Rb} \right)
- \left( \delta \phi_B - \left(\frac{k_\mathrm{eff}}{k_\mathrm{eff}^\mathrm{Rb}}\right) \delta \phi_\mathrm{Rb}' \right)
= - \Delta m\, c^2  \, (\Delta\bar{\tau}_b - \Delta\bar{\tau}_a) / \hbar
= - \Delta m \, \big(1 + \alpha_\text{e-g} \big)\, g \, \Delta z \, (t_\text{f} - t_\text{i}) / \hbar
\label{eq:doubly_differential} , 
\end{equation}
\end{widetext}
where we have used a prime in $\delta \phi_\mathrm{Rb}'$ for the second shot to indicate that it varies from shot to shot due to $\Delta \mathbf{g}$ and the mirror vibrations. The UFF violations from the terms proportional to $\beta_\mathrm{Rb}$ and $\mathbf{g}$, on the other hand, do cancel out in the doubly differential measurement, which constitutes a direct test of UGR with possible violations parametrized by $\alpha_\text{e-g}$.

Note that $\big(k_\mathrm{eff} / k_\mathrm{eff}^\mathrm{Rb} \big) \neq 1$ implies that only common phase-shift contributions in the range $(-\pi, \pi)$ will properly cancel out when subtracting the measured phase shift for the Rb interferometer. This means that other methods, such as vibration isolation or postcorrection with classical inertial sensors, will still be necessary to keep within that range the phase shifts induced by vibration noise and $\Delta \mathbf{g}$. However, the doubly differential measurement in Eq.~\eqref{eq:doubly_differential} helps to relax the requirements on the suppression level attained by those methods by three orders of magnitude.

Moreover, one can guarantee that small fluctuations of $\delta \bar{\varphi}$ and $\delta \bar{\varphi}_\mathrm{Rb}$ cancel out in the differential measurement by generating the chirp rates $\Delta \dot{\omega}$ for both atomic species from the same local oscillator but with a relative rescaling factor $\big(k_\mathrm{eff} / k_\mathrm{eff}^\mathrm{Rb} \big)$.

\subsubsection*{Practical implementation}

Remarkably, the scheme's sensitivity is capable of measuring the gravitational redshift for a delocalized quantum superposition within a parameter regime accessible to current experiments.
Indeed, for Yb atoms (with the clock transition at $\lambda = 578\, \text{nm}$) a vertical separation
of $\Delta z = 1\, \text{cm}$ and a time $(t_\text{f} - t_\text{i}) = 1\, \text{s}$ lead 
to a differential phase shift of $3.5\, \text{mrad}$, %3.55 mrad
as given by Eq.~\eqref{eq:inversions}. Hence, for $N \approx 10^5$ detected atoms and a precision close to the shot-noise limit, the effect can be resolved in a few shots. And even for a phase resolution of $50\, \text{mrad}$, it can be resolved in a few hundred shots. With the third-order Bragg-diffraction pulses employed in Ref.~\cite{plotkin-swing18}, corresponding to the blue and green arrows in Fig.~\ref{fig:transitions}, this $\Delta z$ can be reached in a time $T' = 0.4\, \text{s}$. Larger momentum transfers can be achieved with a sequence of several Bragg pulses \cite{plotkin-swing18}, which enables larger separations $\Delta z$, shorter times $T'$ or both.
For example, just by doubling the total momentum transfer, one can reduce this time to $T' = 0.2\, \text{s}$, so that for a total interferometer time of $2\, \text{s}$, one would still be left with a time $T = 1.6\, \text{s}$ during which the inversion pulses can be applied. This can be exploited to increase the range of values available for $(t_\text{f} - t_\text{i})$, or to check for systematic effects by varying $t_\text{i}$ and $t_\text{f}$ while keeping $(t_\text{f} - t_\text{i})$ fixed.

Furthermore, this simpler atom optics can be applied to bosonic isotopes, such as $^{170}\text{Yb}$ and $^{174}\text{Yb}$, which can be easily cooled down to ultralow temperatures and even reach Bose-Einstein condensation \cite{takasu03,fukuhara07,doerscher13,roy16}. Combined with atomic lensing techniques \cite{muentinga13,kovachy15a}, one can thus achieve very low expansion rates for the atomic cloud, which enable free evolution times of more than $2.8\, \text{s}$. The resulting narrow momentum distribution also guarantees high diffraction efficiencies for the Bragg pulses, even for LMT sequences \cite{chiow11,plotkin-swing18,kovachy15b,gebbe19}.

Finally, for several seconds of interferometer time the effects of rotations and gravity gradients can be quite detrimental. However, they can both be successfully mitigated with compensation techniques which are respectively based on the use of a tip-tilt mirror for retroreflection \cite{lan12,dickerson13} and suitable 
frequency changes of the intermediate laser pulses \cite{roura17a,d_amico17,overstreet18}.

%\begin{figure}[h]
%\begin{center}
%\includegraphics[width=4.5cm]{figures/figure4.pdf}
%\vspace{-0.5ex}
%\end{center}
%\caption{Schematic diagram of the VLBAI facility, a 10-meter atomic fountain capable of simultaneously operating Rb and Yb atom interferometers. The directions of the laser beams employed for the inversion  \comment{(red)} and Bragg diffraction \comment{(green and blue)} pulses are also displayed. The upward propagating beams have been retroreflected by the vibrationally isolated mirror at the bottom.}
%\label{fig:VLBAI}
%\end{figure}

%%%%%
%\section{Discussion}
%%%%%
\subsection*{\large Discussion}

We have proposed an experiment that will be able to measure for the first time arm-dependent gravitational time-dilation effects in a delocalized quantum superposition.
Together with tests of UFF involving quantum superpositions of internal states \cite{rosi17b} or entangled states of two different isotopes \cite{geiger18}, the experiment will push investigations of the equivalence principle further into the quantum regime.
HITec's VLBAI facility in Hannover \cite{hitec} constitutes an ideal
candidate for implementing the proposal. The 10-meter atomic fountain will enable up to $2.8\, \text{s}$ of free evolution and simultaneous operation of Yb and Rb atom interferometers \cite{hartwig15,schlippert20}. Furthermore, its vibration isolation and mitigation system will keep vibration noise below the $10^{-9} g / \sqrt{\text{Hz}}$ level, so that the phase shift induced by any residual vibrations will stay well within the $(-\pi,\pi)$ range, which guarantees its cancellation through 
the differential measurement with the Rb interferometer.
A key advantage of the proposed scheme is the simplicity of the atom optics, involving standard Bragg diffraction, which enables beam-splitter pulses and even LMT sequences with high diffraction efficiencies and very mild requirements on laser power as well as low complexity.

Before concluding, it is worth discussing the comparison with a related proposal in Ref.~\cite{ufrecht20} in order to highlight the crucial 
advantages of the interferometry scheme introduced here, especially concerning its practical implementation.
The measurement put forward there 
involves a variant of the Ramsey-Bord\'e interferometer where the usual diffraction pulses are replaced by sets of pulses mimicking double Raman diffraction \cite{leveque09} between the two clock states, and each based on a sequence of Bragg pulses immediately followed by an inversion pulse.
By subtracting the outcomes for pairs of interferometer shots where the initial internal states are swapped but identical laser pulses are employed, one can be sensitive to a combination of possible violations of UFF and UGR with a single atomic species. In order to disentangle both contributions, it is therefore necessary to perform different sets of measurements separately varying both $T$ and $T'$.
Instead, the differential measurement 
between shots $A$ and~$B$ in Fig.~\ref{fig:inversions}
provides a direct measurement of the gravitational time dilation and possible violations of UGR.

More importantly, in Ref.~\cite{ufrecht20} all Bragg pulses must be capable of efficiently diffracting both internal states. 
In contrast, a decisive advantage of the new scheme presented here is
that no such pulses are needed and even LMT beam splitters with no substantial atom losses can be easily implemented.

Moreover, in contrast to the quantum-clock interferometer in Fig.~\ref{fig:doubly_differential}, the scheme of Ref.~\cite{ufrecht20} is sensitive to vibration noise, which requires long integration times to average it out, and to small changes of the gravitational field from shot to shot. 
Similarly, the differential measurement of the two separate shots described in Fig.~\ref{fig:inversions} is also sensitive to those.
Nevertheless, the new interferometry scheme proposed here overcomes this limitation thanks to the simultaneous operation of an interferometer with Rb atoms that acts as an inertial reference.

%\newpage

\subsection*{\large Acknowledgements}
%\acknowledgements

The authors thank Fabio Di Pumpo, Alexander Friedrich, Enno Giese, Christian Ufrecht
and Wolfgang Schleich for interesting discussions.
This work has been partially supported by the German Aerospace Center (DLR)
with funds provided by the Federal Ministry of Economics and Energy (BMWi)
under Grants No.~50WM1556 (QUANTUS IV) and 50WM1956 (QUANTUS V),
%.
and by the German Research Foundation (DFG) through CRC 1227 (DQmat), projects B07 and B09, and through EXC 2123 (Quantum Frontiers), research units B02 and B05. D.~S. gratefully acknowledges funding by the Federal Ministry of Education and Research (BMBF) through the funding program Photonics Research Germany under contract No.~13N14875.

%%%%%%%%%%%%%%%%%%
\vspace{5.0ex}
%%%%%%%%%%%%%%%%%%
%\vspace{10.0ex}

%%\newpage
%
%%\onecolumngrid
%%\vspace{5.0ex}
%%\twocolumngrid

\subsection*{\large Appendices}

\appendix

\section{Atom interferometry}
\label{app:atom_interferometry}

The state at the first exit port~(I) of the atom interferometer is a quantum superposition of atomic wave packets that have followed the two different arms ($a$ and $b$):
\begin{equation}
| \psi_\text{I} \rangle = \frac{1}{2} \big( e^{i \phi_a} + e^{i \phi_b} \big)\, | \psi_\text{c} \rangle
\label{eq:exit_port1} ,
\end{equation}
where $\phi_a$ and $\phi_b$ are the phases accumulated along each arm. They include the propagation phase, governed by Eqs.~\eqref{eq:prop_phase1}--\eqref{eq:prop_phase2}, as well as the phases associated with the diffraction pulses.
The probability that an atom is detected in the first port is given by
\begin{equation}
\langle \psi_\text{I} | \psi_\text{I} \rangle
= \frac{1}{2} \big(1 + \cos \delta\phi \big)
\label{eq:exit_port2} .
\end{equation}
in terms of the phase shift $\delta\phi = \phi_b - \phi_a$
\footnote{For open interferometers there is an additional contribution corresponding to the so-called separation phase $\delta \phi_\text{sep}$ \cite{roura20a}. However, in this paper we focus on closed interferomters, i.e., with no relative displacement between the interfering wave packets.}.
An analogous result holds for the second exit port~(II), but with the opposite sign in front of the cosine.
Hence, one can infer the phase shift $\delta\phi$ by measuring the fraction of atoms detected in each port.

The phase shift $\delta\phi$ involves contributions of the propagation phases and the phases associated with the diffraction pulses: $\delta\phi = \delta\phi_\text{propag} + \delta\phi_\text{laser}$.
The term $\delta\phi_\text{propag}$ is obtained by calculating the proper time along the central trajectory for each arm using Eq.~\eqref{eq:prop_phase2} and taking into account the changes of internal state, which means that one needs to use the rest mass $m_1 + \Delta m$ for those segments of the central trajectory where the atoms are in the excited state.

On the other hand, $\delta\phi_\text{laser}$ results from adding the phases associated with the diffraction pulses. Neglecting the effects of finite pulse duration, the atomic wave packets acquire a phase $\varepsilon_j \big(\mathbf{k}_\mathrm{eff} \cdot \mathbf{X}(t_j) + \varphi_j \big) + \pi/2$, with $\varepsilon_j = \pm 1$, whenever they are diffracted  by a laser pulse. The index $j$ labels the pulse, $\mathbf{X}(t_j)$ is the central position of the diffracted
atomic wave packet when the pulse is applied, and $\varphi_j$ is a specific phase for each pulse.
For a reversed Ramsey-Bord\'e interferometer $\varepsilon_1 = \varepsilon_3 = 1$ and $\varepsilon_2 = \varepsilon_4 = -1$.
When considering the central trajectories for such an interferometer in a uniform gravitational field, as depicted in Fig.~\ref{fig:inversions}, and adding the contributions from all the laser pulses, the result for $\delta\phi_\text{laser}$ is given by Eq.~\eqref{eq:laser_phase_shift}.

\section{Derivation of the differential phase shift}
\label{app:differential}

In the absence of vibration noise, fluctuations of the injected laser phases and small changes of the gravitational acceleration from shot to shot, the phase-shift contribution $\delta\phi_\text{laser}$ is identical for shots $A$ and $B$ in Fig.~\ref{fig:inversions}. Thus, it cancels out in the differential phase shift $\delta\phi_\mathrm{A} - \delta\phi_\mathrm{B}$, which is entirely determined by the contributions
%$\delta\phi_\text{propag}$
of the propagation phases, specified by Eq.~\eqref{eq:prop_phase1}. More precisely, there will be a non-vanishing contribution from the time interval between the inversion pulses:
\begin{align}
\delta \phi_A - \delta \phi_B &= 
- (m_1 + \Delta m)\, c^2  \, (\Delta\tau_b - \Delta\tau_a) / \hbar
\nonumber \\
&\quad + m_1 \, c^2  \, (\Delta\tau_b - \Delta\tau_a) / \hbar
\nonumber \\
&= - \Delta m \, c^2  \, (\Delta\tau_b - \Delta\tau_a) / \hbar
\label{eq:diff_phase_shift1}\, ,
\end{align}
where $\Delta\tau_a$ and $\Delta\tau_b$ are the proper times on both arms for the segments of the central trajectories where the atoms are in the excited state. The non-vanishing proper-time difference $\Delta\tau_b - \Delta\tau_a$ is due to the gravitational time dilation, and for non-relativistic velocities and weak fields it can be calculated through Eq.~\eqref{eq:prop_phase2}:
\begin{align}
\delta \phi_A - \delta \phi_B &= 
- \Delta m\, c^2  \, (\Delta\tau_b - \Delta\tau_a) / \hbar
\nonumber \\
%= \frac{\Delta E}{2 \hbar} \, (\Delta\tau_b - \Delta\tau_a)
&= - \Delta m \, g \, \Delta z \, (t_\text{f} - t_\text{i}) / \hbar
\label{eq:diff_phase_shift2}\, .
\end{align}

Although the atoms experience in the laboratory frame no net momentum transfer from the inversion pulses, the change of rest mass leads to a slight change of the central velocity, $\Delta \mathbf{v} = - (\Delta m / m)\, \mathbf{v}$, for atoms that are not initially at rest.
However, this residual recoil does not alter the result for $\delta \phi_A - \delta \phi_B$. Indeed, the velocity change is the same for both arms and the arm separation is still $\Delta z$. Hence, the net contributions of the kinetic and potential terms to the phase difference between the two arms before the last pair of diffraction pulses remains unchanged.
On the other hand, the phase difference from the kinetic terms is slightly modified between the last two diffraction pulses, but so is the contribution to $\delta\phi_\text{laser}$ from those two pulses and both changes exactly cancel out, as explained in Ref.~\cite{roura20a}.

In reality, vibration noise and small changes of the gravitational acceleration from shot to shot prevent $\delta\phi_\text{laser}$ from canceling out in the differential phase shift $\delta \phi_A - \delta \phi_B$. Nevertheless, as explained in the main text, this drawback can be overcome by simultaneously operating a Rb interferometer that acts as an inertial reference.

%\vspace{0.5ex}

%\section{Derivation of the 
\section{Modifications due to dilaton models}
\label{app:differential_dilaton}

Dilaton models provide a consistent framework for parametrizing violations of the equivalence principle \cite{damour12,damour10a}. In the regime of non-relativistic velocities and weak fields their effect on test particles amounts to replacing the potential term in Eq.~\eqref{eq:prop_phase2} with $m_n (1 + \beta_n)\, U (t',\mathbf{X})$. For a uniform field this corresponds to the substitution $\mathbf{g} \to (1 + \beta_n)\,\mathbf{g}$ in the calculation of the differential phase shift:
\begin{align}
\delta \phi_A - \delta \phi_B &= 
- (m_1 + \Delta m)\, (1+ \beta_2)\, g \, \Delta z \, (t_\text{f} - t_\text{i}) / \hbar
\nonumber \\
&\quad + m_1 \, (1+ \beta_1)\, g \, \Delta z \, (t_\text{f} - t_\text{i}) / \hbar
\nonumber \\
&\approx - \Delta m \, \big(1 + \alpha_\text{e-g} \big)\, g \, \Delta z \, (t_\text{f} - t_\text{i}) / \hbar
\label{eq:diff_phase_shift3}\, ,
\end{align}
where a term proportional to $\Delta m \, \beta_2$ has been neglected in the last equality and the parameter
\begin{equation}
\alpha_\text{e-g} = \frac{m_1}{\Delta m} \big( \beta_2 - \beta_1 \big)
\label{eq:alpha2} ,
\end{equation}
has been introduced.

Note that $\beta_2 \neq \beta_1$ implies a slightly different acceleration for the excited state, which leads to a small modification of the central trajectories for shot $A$ compared to shot~$B$. However, the deviation is the same for both arms and $\delta \phi_A - \delta \phi_B$ remains unchanged for the same %analogous
reasons as for the residual recoil discussed in Appendix~\ref{app:differential}.

\vspace*{5.0ex}

%The next line should be commented out for the titles to appear.
%\bibliographystyle{apsrev4-1}
\bibliography{literature5}

\begin{thebibliography}{80}%
\makeatletter
\providecommand \@ifxundefined [1]{%
 \@ifx{#1\undefined}
}%
\providecommand \@ifnum [1]{%
 \ifnum #1\expandafter \@firstoftwo
 \else \expandafter \@secondoftwo
 \fi
}%
\providecommand \@ifx [1]{%
 \ifx #1\expandafter \@firstoftwo
 \else \expandafter \@secondoftwo
 \fi
}%
\providecommand \natexlab [1]{#1}%
\providecommand \enquote  [1]{``#1''}%
\providecommand \bibnamefont  [1]{#1}%
\providecommand \bibfnamefont [1]{#1}%
\providecommand \citenamefont [1]{#1}%
\providecommand \href@noop [0]{\@secondoftwo}%
\providecommand \href [0]{\begingroup \@sanitize@url \@href}%
\providecommand \@href[1]{\@@startlink{#1}\@@href}%
\providecommand \@@href[1]{\endgroup#1\@@endlink}%
\providecommand \@sanitize@url [0]{\catcode `\\12\catcode `\$12\catcode
  `\&12\catcode `\#12\catcode `\^12\catcode `\_12\catcode `\%12\relax}%
\providecommand \@@startlink[1]{}%
\providecommand \@@endlink[0]{}%
\providecommand \url  [0]{\begingroup\@sanitize@url \@url }%
\providecommand \@url [1]{\endgroup\@href {#1}{\urlprefix }}%
\providecommand \urlprefix  [0]{URL }%
\providecommand \Eprint [0]{\href }%
\providecommand \doibase [0]{https://doi.org/}%
\providecommand \selectlanguage [0]{\@gobble}%
\providecommand \bibinfo  [0]{\@secondoftwo}%
\providecommand \bibfield  [0]{\@secondoftwo}%
\providecommand \translation [1]{[#1]}%
\providecommand \BibitemOpen [0]{}%
\providecommand \bibitemStop [0]{}%
\providecommand \bibitemNoStop [0]{.\EOS\space}%
\providecommand \EOS [0]{\spacefactor3000\relax}%
\providecommand \BibitemShut  [1]{\csname bibitem#1\endcsname}%
\let\auto@bib@innerbib\@empty
%</preamble>
\bibitem [{\citenamefont {Kasevich}\ and\ \citenamefont
  {Chu}(1991)}]{kasevich91}%
  \BibitemOpen
  \bibfield  {author} {\bibinfo {author} {\bibfnamefont {M.}~\bibnamefont
  {Kasevich}}\ and\ \bibinfo {author} {\bibfnamefont {S.}~\bibnamefont {Chu}},\
  }\bibfield  {title} {\bibinfo {title} {Atomic interferometry using stimulated
  {Raman} transitions},\ }\href {https://doi.org/10.1103/PhysRevLett.67.181}
  {\bibfield  {journal} {\bibinfo  {journal} {Phys. Rev. Lett.}\ }\textbf
  {\bibinfo {volume} {67}},\ \bibinfo {pages} {181} (\bibinfo {year}
  {1991})}\BibitemShut {NoStop}%
\bibitem [{\citenamefont {Kleinert}\ \emph {et~al.}(2015)\citenamefont
  {Kleinert}, \citenamefont {Kajari}, \citenamefont {Roura},\ and\
  \citenamefont {Schleich}}]{kleinert15}%
  \BibitemOpen
  \bibfield  {author} {\bibinfo {author} {\bibfnamefont {S.}~\bibnamefont
  {Kleinert}}, \bibinfo {author} {\bibfnamefont {E.}~\bibnamefont {Kajari}},
  \bibinfo {author} {\bibfnamefont {A.}~\bibnamefont {Roura}},\ and\ \bibinfo
  {author} {\bibfnamefont {W.~P.}\ \bibnamefont {Schleich}},\ }\bibfield
  {title} {\bibinfo {title} {Representation-free description of light-pulse
  atom interferometry including noninertial effects},\ }\href
  {http://dx.doi.org/10.1016/j.physrep.2015.09.004} {\bibfield  {journal}
  {\bibinfo  {journal} {Phys. Reports}\ }\textbf {\bibinfo {volume} {605}},\
  \bibinfo {pages} {1} (\bibinfo {year} {2015})}\BibitemShut {NoStop}%
\bibitem [{\citenamefont {Kovachy}\ \emph
  {et~al.}(2015{\natexlab{a}})\citenamefont {Kovachy}, \citenamefont
  {Asenbaum}, \citenamefont {Overstreet}, \citenamefont {Donnelly},
  \citenamefont {Dickerson}, \citenamefont {Sugarbaker}, \citenamefont
  {Hogan},\ and\ \citenamefont {Kasevich}}]{kovachy15b}%
  \BibitemOpen
  \bibfield  {author} {\bibinfo {author} {\bibfnamefont {T.}~\bibnamefont
  {Kovachy}}, \bibinfo {author} {\bibfnamefont {P.}~\bibnamefont {Asenbaum}},
  \bibinfo {author} {\bibfnamefont {C.}~\bibnamefont {Overstreet}}, \bibinfo
  {author} {\bibfnamefont {C.~A.}\ \bibnamefont {Donnelly}}, \bibinfo {author}
  {\bibfnamefont {S.~M.}\ \bibnamefont {Dickerson}}, \bibinfo {author}
  {\bibfnamefont {A.}~\bibnamefont {Sugarbaker}}, \bibinfo {author}
  {\bibfnamefont {J.~M.}\ \bibnamefont {Hogan}},\ and\ \bibinfo {author}
  {\bibfnamefont {M.~A.}\ \bibnamefont {Kasevich}},\ }\bibfield  {title}
  {\bibinfo {title} {Quantum superposition at the half-metre scale},\
  }\href@noop {} {\bibfield  {journal} {\bibinfo  {journal} {Nature}\ }\textbf
  {\bibinfo {volume} {528}},\ \bibinfo {pages} {530} (\bibinfo {year}
  {2015}{\natexlab{a}})}\BibitemShut {NoStop}%
\bibitem [{\citenamefont {Peters}\ \emph {et~al.}(1999)\citenamefont {Peters},
  \citenamefont {Chung},\ and\ \citenamefont {Chu}}]{peters99}%
  \BibitemOpen
  \bibfield  {author} {\bibinfo {author} {\bibfnamefont {A.}~\bibnamefont
  {Peters}}, \bibinfo {author} {\bibfnamefont {K.~Y.}\ \bibnamefont {Chung}},\
  and\ \bibinfo {author} {\bibfnamefont {S.}~\bibnamefont {Chu}},\ }\bibfield
  {title} {\bibinfo {title} {Measurement of gravitational acceleration by
  dropping atoms},\ }\href {http://dx.doi.org/10.1038/23655} {\bibfield
  {journal} {\bibinfo  {journal} {Nature}\ }\textbf {\bibinfo {volume} {400}},\
  \bibinfo {pages} {849} (\bibinfo {year} {1999})}\BibitemShut {NoStop}%
\bibitem [{\citenamefont {Savoie}\ \emph {et~al.}(2018)\citenamefont {Savoie},
  \citenamefont {Altorio}, \citenamefont {Fang}, \citenamefont {Sidorenkov},
  \citenamefont {Geiger},\ and\ \citenamefont {Landragin}}]{savoie18}%
  \BibitemOpen
  \bibfield  {author} {\bibinfo {author} {\bibfnamefont {D.}~\bibnamefont
  {Savoie}}, \bibinfo {author} {\bibfnamefont {M.}~\bibnamefont {Altorio}},
  \bibinfo {author} {\bibfnamefont {B.}~\bibnamefont {Fang}}, \bibinfo {author}
  {\bibfnamefont {L.~A.}\ \bibnamefont {Sidorenkov}}, \bibinfo {author}
  {\bibfnamefont {R.}~\bibnamefont {Geiger}},\ and\ \bibinfo {author}
  {\bibfnamefont {A.}~\bibnamefont {Landragin}},\ }\bibfield  {title} {\bibinfo
  {title} {Interleaved atom interferometry for high-sensitivity inertial
  measurements},\ }\bibfield  {journal} {\bibinfo  {journal} {Science Adv.}\
  }\textbf {\bibinfo {volume} {4}},\ \href
  {https://doi.org/10.1126/sciadv.aau7948} {10.1126/sciadv.aau7948} (\bibinfo
  {year} {2018})\BibitemShut {NoStop}%
\bibitem [{\citenamefont {Bongs}\ \emph {et~al.}(2019)\citenamefont {Bongs},
  \citenamefont {Holynski}, \citenamefont {Vovrosh}, \citenamefont {Bouyer},
  \citenamefont {Condon}, \citenamefont {Rasel}, \citenamefont {Schubert},
  \citenamefont {Schleich},\ and\ \citenamefont {Roura}}]{bongs19}%
  \BibitemOpen
  \bibfield  {author} {\bibinfo {author} {\bibfnamefont {K.}~\bibnamefont
  {Bongs}}, \bibinfo {author} {\bibfnamefont {M.}~\bibnamefont {Holynski}},
  \bibinfo {author} {\bibfnamefont {J.}~\bibnamefont {Vovrosh}}, \bibinfo
  {author} {\bibfnamefont {P.}~\bibnamefont {Bouyer}}, \bibinfo {author}
  {\bibfnamefont {G.}~\bibnamefont {Condon}}, \bibinfo {author} {\bibfnamefont
  {E.}~\bibnamefont {Rasel}}, \bibinfo {author} {\bibfnamefont
  {C.}~\bibnamefont {Schubert}}, \bibinfo {author} {\bibfnamefont {W.~P.}\
  \bibnamefont {Schleich}},\ and\ \bibinfo {author} {\bibfnamefont
  {A.}~\bibnamefont {Roura}},\ }\bibfield  {title} {\bibinfo {title} {Taking
  atom interferometric quantum sensors from the laboratory to real-world
  applications},\ }\href {https://doi.org/10.1038/s42254-019-0117-4} {\bibfield
   {journal} {\bibinfo  {journal} {Nature Rev. Phys.}\ }\textbf {\bibinfo
  {volume} {1}},\ \bibinfo {pages} {731} (\bibinfo {year} {2019})}\BibitemShut
  {NoStop}%
\bibitem [{\citenamefont {{Rosi}}\ \emph {et~al.}(2014)\citenamefont {{Rosi}},
  \citenamefont {{Sorrentino}}, \citenamefont {{Cacciapuoti}}, \citenamefont
  {{Prevedelli}},\ and\ \citenamefont {{Tino}}}]{rosi14}%
  \BibitemOpen
  \bibfield  {author} {\bibinfo {author} {\bibfnamefont {G.}~\bibnamefont
  {{Rosi}}}, \bibinfo {author} {\bibfnamefont {F.}~\bibnamefont
  {{Sorrentino}}}, \bibinfo {author} {\bibfnamefont {L.}~\bibnamefont
  {{Cacciapuoti}}}, \bibinfo {author} {\bibfnamefont {M.}~\bibnamefont
  {{Prevedelli}}},\ and\ \bibinfo {author} {\bibfnamefont {G.~M.}\ \bibnamefont
  {{Tino}}},\ }\bibfield  {title} {\bibinfo {title} {{Precision measurement of
  the Newtonian gravitational constant using cold atoms}},\ }\href
  {https://doi.org/10.1038/nature13433} {\bibfield  {journal} {\bibinfo
  {journal} {Nature}\ }\textbf {\bibinfo {volume} {510}},\ \bibinfo {pages}
  {518} (\bibinfo {year} {2014})}\BibitemShut {NoStop}%
\bibitem [{\citenamefont {Bouchendira}\ \emph {et~al.}(2011)\citenamefont
  {Bouchendira}, \citenamefont {Clad\'{e}}, \citenamefont
  {Guellati-Kh\'{e}lifa}, \citenamefont {Nez},\ and\ \citenamefont
  {Biraben}}]{bouchendira11}%
  \BibitemOpen
  \bibfield  {author} {\bibinfo {author} {\bibfnamefont {R.}~\bibnamefont
  {Bouchendira}}, \bibinfo {author} {\bibfnamefont {P.}~\bibnamefont
  {Clad\'{e}}}, \bibinfo {author} {\bibfnamefont {S.}~\bibnamefont
  {Guellati-Kh\'{e}lifa}}, \bibinfo {author} {\bibfnamefont {F.}~\bibnamefont
  {Nez}},\ and\ \bibinfo {author} {\bibfnamefont {F.}~\bibnamefont {Biraben}},\
  }\bibfield  {title} {\bibinfo {title} {New determination of the fine
  structure constant and test of the quantum electrodynamics},\ }\href
  {https://doi.org/10.1103/PhysRevLett.106.080801} {\bibfield  {journal}
  {\bibinfo  {journal} {Phys. Rev. Lett.}\ }\textbf {\bibinfo {volume} {106}},\
  \bibinfo {pages} {080801} (\bibinfo {year} {2011})}\BibitemShut {NoStop}%
\bibitem [{\citenamefont {Parker}\ \emph {et~al.}(2018)\citenamefont {Parker},
  \citenamefont {Yu}, \citenamefont {Zhong}, \citenamefont {Estey},\ and\
  \citenamefont {M{\"u}ller}}]{parker18}%
  \BibitemOpen
  \bibfield  {author} {\bibinfo {author} {\bibfnamefont {R.~H.}\ \bibnamefont
  {Parker}}, \bibinfo {author} {\bibfnamefont {C.}~\bibnamefont {Yu}}, \bibinfo
  {author} {\bibfnamefont {W.}~\bibnamefont {Zhong}}, \bibinfo {author}
  {\bibfnamefont {B.}~\bibnamefont {Estey}},\ and\ \bibinfo {author}
  {\bibfnamefont {H.}~\bibnamefont {M{\"u}ller}},\ }\bibfield  {title}
  {\bibinfo {title} {Measurement of the fine-structure constant as a test of
  the {Standard} {Model}},\ }\href@noop {} {\bibfield  {journal} {\bibinfo
  {journal} {Science}\ }\textbf {\bibinfo {volume} {360}},\ \bibinfo {pages}
  {191} (\bibinfo {year} {2018})}\BibitemShut {NoStop}%
\bibitem [{\citenamefont {Schlippert}\ \emph {et~al.}(2014)\citenamefont
  {Schlippert}, \citenamefont {Hartwig}, \citenamefont {Albers}, \citenamefont
  {Richardson}, \citenamefont {Schubert}, \citenamefont {Roura}, \citenamefont
  {Schleich}, \citenamefont {Ertmer},\ and\ \citenamefont
  {Rasel}}]{schlippert14}%
  \BibitemOpen
  \bibfield  {author} {\bibinfo {author} {\bibfnamefont {D.}~\bibnamefont
  {Schlippert}}, \bibinfo {author} {\bibfnamefont {J.}~\bibnamefont {Hartwig}},
  \bibinfo {author} {\bibfnamefont {H.}~\bibnamefont {Albers}}, \bibinfo
  {author} {\bibfnamefont {L.~L.}\ \bibnamefont {Richardson}}, \bibinfo
  {author} {\bibfnamefont {C.}~\bibnamefont {Schubert}}, \bibinfo {author}
  {\bibfnamefont {A.}~\bibnamefont {Roura}}, \bibinfo {author} {\bibfnamefont
  {W.~P.}\ \bibnamefont {Schleich}}, \bibinfo {author} {\bibfnamefont
  {W.}~\bibnamefont {Ertmer}},\ and\ \bibinfo {author} {\bibfnamefont {E.~M.}\
  \bibnamefont {Rasel}},\ }\bibfield  {title} {\bibinfo {title} {Quantum test
  of the universality of free fall},\ }\href
  {https://doi.org/10.1103/PhysRevLett.112.203002} {\bibfield  {journal}
  {\bibinfo  {journal} {Phys. Rev. Lett.}\ }\textbf {\bibinfo {volume} {112}},\
  \bibinfo {pages} {203002} (\bibinfo {year} {2014})}\BibitemShut {NoStop}%
\bibitem [{\citenamefont {Zhou}\ \emph {et~al.}(2015)\citenamefont {Zhou},
  \citenamefont {Long}, \citenamefont {Tang}, \citenamefont {Chen},
  \citenamefont {Gao}, \citenamefont {Peng}, \citenamefont {Duan},
  \citenamefont {Zhong}, \citenamefont {Xiong}, \citenamefont {Wang},
  \citenamefont {Zhang},\ and\ \citenamefont {Zhan}}]{zhou15}%
  \BibitemOpen
  \bibfield  {author} {\bibinfo {author} {\bibfnamefont {L.}~\bibnamefont
  {Zhou}}, \bibinfo {author} {\bibfnamefont {S.}~\bibnamefont {Long}}, \bibinfo
  {author} {\bibfnamefont {B.}~\bibnamefont {Tang}}, \bibinfo {author}
  {\bibfnamefont {X.}~\bibnamefont {Chen}}, \bibinfo {author} {\bibfnamefont
  {F.}~\bibnamefont {Gao}}, \bibinfo {author} {\bibfnamefont {W.}~\bibnamefont
  {Peng}}, \bibinfo {author} {\bibfnamefont {W.}~\bibnamefont {Duan}}, \bibinfo
  {author} {\bibfnamefont {J.}~\bibnamefont {Zhong}}, \bibinfo {author}
  {\bibfnamefont {Z.}~\bibnamefont {Xiong}}, \bibinfo {author} {\bibfnamefont
  {J.}~\bibnamefont {Wang}}, \bibinfo {author} {\bibfnamefont {Y.}~\bibnamefont
  {Zhang}},\ and\ \bibinfo {author} {\bibfnamefont {M.}~\bibnamefont {Zhan}},\
  }\bibfield  {title} {\bibinfo {title} {Test of equivalence principle at
  $1{0}^{-8}$ level by a dual-species double-diffraction {Raman} atom
  interferometer},\ }\href {https://doi.org/10.1103/PhysRevLett.115.013004}
  {\bibfield  {journal} {\bibinfo  {journal} {Phys. Rev. Lett.}\ }\textbf
  {\bibinfo {volume} {115}},\ \bibinfo {pages} {013004} (\bibinfo {year}
  {2015})}\BibitemShut {NoStop}%
\bibitem [{\citenamefont {Rosi}\ \emph {et~al.}(2017)\citenamefont {Rosi},
  \citenamefont {D'Amico}, \citenamefont {Cacciapuoti}, \citenamefont
  {Sorrentino}, \citenamefont {Prevedelli}, \citenamefont {Zych}, \citenamefont
  {Brukner},\ and\ \citenamefont {Tino}}]{rosi17b}%
  \BibitemOpen
  \bibfield  {author} {\bibinfo {author} {\bibfnamefont {G.}~\bibnamefont
  {Rosi}}, \bibinfo {author} {\bibfnamefont {G.}~\bibnamefont {D'Amico}},
  \bibinfo {author} {\bibfnamefont {L.}~\bibnamefont {Cacciapuoti}}, \bibinfo
  {author} {\bibfnamefont {F.}~\bibnamefont {Sorrentino}}, \bibinfo {author}
  {\bibfnamefont {M.}~\bibnamefont {Prevedelli}}, \bibinfo {author}
  {\bibfnamefont {M.}~\bibnamefont {Zych}}, \bibinfo {author} {\bibfnamefont
  {{\v{C}}.}~\bibnamefont {Brukner}},\ and\ \bibinfo {author} {\bibfnamefont
  {G.}~\bibnamefont {Tino}},\ }\bibfield  {title} {\bibinfo {title} {Quantum
  test of the equivalence principle for atoms in coherent superposition of
  internal energy states},\ }\href {https://doi.org/10.1038/ncomms15529}
  {\bibfield  {journal} {\bibinfo  {journal} {Nat. Comm.}\ }\textbf {\bibinfo
  {volume} {8}},\ \bibinfo {pages} {15529} (\bibinfo {year}
  {2017})}\BibitemShut {NoStop}%
\bibitem [{\citenamefont {Asenbaum}\ \emph {et~al.}()\citenamefont {Asenbaum},
  \citenamefont {Overstreet}, \citenamefont {Kim}, \citenamefont {Curti},\ and\
  \citenamefont {Kasevich}}]{asenbaum20}%
  \BibitemOpen
  \bibfield  {author} {\bibinfo {author} {\bibfnamefont {P.}~\bibnamefont
  {Asenbaum}}, \bibinfo {author} {\bibfnamefont {C.}~\bibnamefont
  {Overstreet}}, \bibinfo {author} {\bibfnamefont {M.}~\bibnamefont {Kim}},
  \bibinfo {author} {\bibfnamefont {J.}~\bibnamefont {Curti}},\ and\ \bibinfo
  {author} {\bibfnamefont {M.~A.}\ \bibnamefont {Kasevich}},\ }\bibfield
  {title} {\bibinfo {title} {Atom-interferometric test of the equivalence
  principle at the $10^{-12}$ level},\ }\href@noop {} {\bibinfo  {journal}
  {\texttt{arXiv:2005.11624}}\ }\BibitemShut {NoStop}%
\bibitem [{\citenamefont {Hamilton}\ \emph {et~al.}(2015)\citenamefont
  {Hamilton}, \citenamefont {Jaffe}, \citenamefont {Haslinger}, \citenamefont
  {Simmons}, \citenamefont {M{\"u}ller},\ and\ \citenamefont
  {Khoury}}]{hamilton15b}%
  \BibitemOpen
\bibfield  {journal} {  }\bibfield  {author} {\bibinfo {author} {\bibfnamefont
  {P.}~\bibnamefont {Hamilton}}, \bibinfo {author} {\bibfnamefont
  {M.}~\bibnamefont {Jaffe}}, \bibinfo {author} {\bibfnamefont
  {P.}~\bibnamefont {Haslinger}}, \bibinfo {author} {\bibfnamefont
  {Q.}~\bibnamefont {Simmons}}, \bibinfo {author} {\bibfnamefont
  {H.}~\bibnamefont {M{\"u}ller}},\ and\ \bibinfo {author} {\bibfnamefont
  {J.}~\bibnamefont {Khoury}},\ }\bibfield  {title} {\bibinfo {title}
  {Atom-interferometry constraints on dark energy},\ }\href
  {https://doi.org/10.1126/science.aaa8883} {\bibfield  {journal} {\bibinfo
  {journal} {Science}\ }\textbf {\bibinfo {volume} {349}},\ \bibinfo {pages}
  {849} (\bibinfo {year} {2015})}\BibitemShut {NoStop}%
\bibitem [{\citenamefont {Jaffe}\ \emph {et~al.}(2017)\citenamefont {Jaffe},
  \citenamefont {Haslinger}, \citenamefont {Xu}, \citenamefont {Hamilton},
  \citenamefont {Upadhye}, \citenamefont {Elder}, \citenamefont {Khoury},\ and\
  \citenamefont {M{\"u}ller}}]{jaffe17}%
  \BibitemOpen
  \bibfield  {author} {\bibinfo {author} {\bibfnamefont {M.}~\bibnamefont
  {Jaffe}}, \bibinfo {author} {\bibfnamefont {P.}~\bibnamefont {Haslinger}},
  \bibinfo {author} {\bibfnamefont {V.}~\bibnamefont {Xu}}, \bibinfo {author}
  {\bibfnamefont {P.}~\bibnamefont {Hamilton}}, \bibinfo {author}
  {\bibfnamefont {A.}~\bibnamefont {Upadhye}}, \bibinfo {author} {\bibfnamefont
  {B.}~\bibnamefont {Elder}}, \bibinfo {author} {\bibfnamefont
  {J.}~\bibnamefont {Khoury}},\ and\ \bibinfo {author} {\bibfnamefont
  {H.}~\bibnamefont {M{\"u}ller}},\ }\bibfield  {title} {\bibinfo {title}
  {Testing sub-gravitational forces on atoms from a miniature in-vacuum source
  mass},\ }\href {https://doi.org/10.1038/nphys4189} {\bibfield  {journal}
  {\bibinfo  {journal} {Nature Phys.}\ }\textbf {\bibinfo {volume} {13}},\
  \bibinfo {pages} {938} (\bibinfo {year} {2017})}\BibitemShut {NoStop}%
\bibitem [{\citenamefont {Sabulsky}\ \emph {et~al.}(2019)\citenamefont
  {Sabulsky}, \citenamefont {Dutta}, \citenamefont {Hinds}, \citenamefont
  {Elder}, \citenamefont {Burrage},\ and\ \citenamefont
  {Copeland}}]{sabulsky19}%
  \BibitemOpen
  \bibfield  {author} {\bibinfo {author} {\bibfnamefont {D.~O.}\ \bibnamefont
  {Sabulsky}}, \bibinfo {author} {\bibfnamefont {I.}~\bibnamefont {Dutta}},
  \bibinfo {author} {\bibfnamefont {E.~A.}\ \bibnamefont {Hinds}}, \bibinfo
  {author} {\bibfnamefont {B.}~\bibnamefont {Elder}}, \bibinfo {author}
  {\bibfnamefont {C.}~\bibnamefont {Burrage}},\ and\ \bibinfo {author}
  {\bibfnamefont {E.~J.}\ \bibnamefont {Copeland}},\ }\bibfield  {title}
  {\bibinfo {title} {Experiment to detect dark energy forces using atom
  interferometry},\ }\href {https://doi.org/10.1103/PhysRevLett.123.061102}
  {\bibfield  {journal} {\bibinfo  {journal} {Phys. Rev. Lett.}\ }\textbf
  {\bibinfo {volume} {123}},\ \bibinfo {pages} {061102} (\bibinfo {year}
  {2019})}\BibitemShut {NoStop}%
\bibitem [{\citenamefont {Bloom}\ \emph {et~al.}(2014)\citenamefont {Bloom},
  \citenamefont {Nicholson}, \citenamefont {Williams}, \citenamefont
  {Campbell}, \citenamefont {Bishof}, \citenamefont {Zhang}, \citenamefont
  {Zhang}, \citenamefont {Bromley},\ and\ \citenamefont {Ye}}]{bloom14}%
  \BibitemOpen
  \bibfield  {author} {\bibinfo {author} {\bibfnamefont {B.~J.}\ \bibnamefont
  {Bloom}}, \bibinfo {author} {\bibfnamefont {T.~L.}\ \bibnamefont
  {Nicholson}}, \bibinfo {author} {\bibfnamefont {J.~R.}\ \bibnamefont
  {Williams}}, \bibinfo {author} {\bibfnamefont {S.~L.}\ \bibnamefont
  {Campbell}}, \bibinfo {author} {\bibfnamefont {M.}~\bibnamefont {Bishof}},
  \bibinfo {author} {\bibfnamefont {X.}~\bibnamefont {Zhang}}, \bibinfo
  {author} {\bibfnamefont {W.}~\bibnamefont {Zhang}}, \bibinfo {author}
  {\bibfnamefont {S.~L.}\ \bibnamefont {Bromley}},\ and\ \bibinfo {author}
  {\bibfnamefont {J.}~\bibnamefont {Ye}},\ }\bibfield  {title} {\bibinfo
  {title} {An optical lattice clock with accuracy and stability at the
  $10^{-18}$ level},\ }\href {https://doi.org/10.1038/nature12941} {\bibfield
  {journal} {\bibinfo  {journal} {Nature}\ }\textbf {\bibinfo {volume} {506}},\
  \bibinfo {pages} {71} (\bibinfo {year} {2014})}\BibitemShut {NoStop}%
\bibitem [{\citenamefont {Campbell}\ \emph {et~al.}(2017)\citenamefont
  {Campbell}, \citenamefont {Hutson}, \citenamefont {Marti}, \citenamefont
  {Goban}, \citenamefont {Darkwah~Oppong}, \citenamefont {McNally},
  \citenamefont {Sonderhouse}, \citenamefont {Robinson}, \citenamefont {Zhang},
  \citenamefont {Bloom},\ and\ \citenamefont {Ye}}]{campbell17}%
  \BibitemOpen
  \bibfield  {author} {\bibinfo {author} {\bibfnamefont {S.~L.}\ \bibnamefont
  {Campbell}}, \bibinfo {author} {\bibfnamefont {R.~B.}\ \bibnamefont
  {Hutson}}, \bibinfo {author} {\bibfnamefont {G.~E.}\ \bibnamefont {Marti}},
  \bibinfo {author} {\bibfnamefont {A.}~\bibnamefont {Goban}}, \bibinfo
  {author} {\bibfnamefont {N.}~\bibnamefont {Darkwah~Oppong}}, \bibinfo
  {author} {\bibfnamefont {R.~L.}\ \bibnamefont {McNally}}, \bibinfo {author}
  {\bibfnamefont {L.}~\bibnamefont {Sonderhouse}}, \bibinfo {author}
  {\bibfnamefont {J.~M.}\ \bibnamefont {Robinson}}, \bibinfo {author}
  {\bibfnamefont {W.}~\bibnamefont {Zhang}}, \bibinfo {author} {\bibfnamefont
  {B.~J.}\ \bibnamefont {Bloom}},\ and\ \bibinfo {author} {\bibfnamefont
  {J.}~\bibnamefont {Ye}},\ }\bibfield  {title} {\bibinfo {title} {A
  {Fermi}-degenerate three-dimensional optical lattice clock},\ }\href
  {https://doi.org/10.1126/science.aam5538} {\bibfield  {journal} {\bibinfo
  {journal} {Science}\ }\textbf {\bibinfo {volume} {358}},\ \bibinfo {pages}
  {90} (\bibinfo {year} {2017})}\BibitemShut {NoStop}%
\bibitem [{\citenamefont {Ludlow}\ \emph {et~al.}(2015)\citenamefont {Ludlow},
  \citenamefont {Boyd}, \citenamefont {Ye}, \citenamefont {Peik},\ and\
  \citenamefont {Schmidt}}]{ludlow15}%
  \BibitemOpen
  \bibfield  {author} {\bibinfo {author} {\bibfnamefont {A.~D.}\ \bibnamefont
  {Ludlow}}, \bibinfo {author} {\bibfnamefont {M.~M.}\ \bibnamefont {Boyd}},
  \bibinfo {author} {\bibfnamefont {J.}~\bibnamefont {Ye}}, \bibinfo {author}
  {\bibfnamefont {E.}~\bibnamefont {Peik}},\ and\ \bibinfo {author}
  {\bibfnamefont {P.~O.}\ \bibnamefont {Schmidt}},\ }\bibfield  {title}
  {\bibinfo {title} {Optical atomic clocks},\ }\href@noop {} {\bibfield
  {journal} {\bibinfo  {journal} {Rev. Mod. Phys.}\ }\textbf {\bibinfo {volume}
  {87}},\ \bibinfo {pages} {637} (\bibinfo {year} {2015})}\BibitemShut
  {NoStop}%
\bibitem [{\citenamefont {Hees}\ \emph {et~al.}(2016)\citenamefont {Hees},
  \citenamefont {Gu\'ena}, \citenamefont {Abgrall}, \citenamefont {Bize},\ and\
  \citenamefont {Wolf}}]{hees16}%
  \BibitemOpen
  \bibfield  {author} {\bibinfo {author} {\bibfnamefont {A.}~\bibnamefont
  {Hees}}, \bibinfo {author} {\bibfnamefont {J.}~\bibnamefont {Gu\'ena}},
  \bibinfo {author} {\bibfnamefont {M.}~\bibnamefont {Abgrall}}, \bibinfo
  {author} {\bibfnamefont {S.}~\bibnamefont {Bize}},\ and\ \bibinfo {author}
  {\bibfnamefont {P.}~\bibnamefont {Wolf}},\ }\bibfield  {title} {\bibinfo
  {title} {Searching for an oscillating massive scalar field as a dark matter
  candidate using atomic hyperfine frequency comparisons},\ }\href
  {https://doi.org/10.1103/PhysRevLett.117.061301} {\bibfield  {journal}
  {\bibinfo  {journal} {Phys. Rev. Lett.}\ }\textbf {\bibinfo {volume} {117}},\
  \bibinfo {pages} {061301} (\bibinfo {year} {2016})}\BibitemShut {NoStop}%
\bibitem [{\citenamefont {Matveev}\ \emph {et~al.}(2013)\citenamefont
  {Matveev}, \citenamefont {Parthey}, \citenamefont {Predehl}, \citenamefont
  {Alnis}, \citenamefont {Beyer}, \citenamefont {Holzwarth}, \citenamefont
  {Udem}, \citenamefont {Wilken}, \citenamefont {Kolachevsky}, \citenamefont
  {Abgrall}, \citenamefont {Rovera}, \citenamefont {Salomon}, \citenamefont
  {Laurent}, \citenamefont {Grosche}, \citenamefont {Terra}, \citenamefont
  {Legero}, \citenamefont {Schnatz}, \citenamefont {Weyers}, \citenamefont
  {Altschul},\ and\ \citenamefont {H\"ansch}}]{matveev13}%
  \BibitemOpen
  \bibfield  {author} {\bibinfo {author} {\bibfnamefont {A.}~\bibnamefont
  {Matveev}}, \bibinfo {author} {\bibfnamefont {C.~G.}\ \bibnamefont
  {Parthey}}, \bibinfo {author} {\bibfnamefont {K.}~\bibnamefont {Predehl}},
  \bibinfo {author} {\bibfnamefont {J.}~\bibnamefont {Alnis}}, \bibinfo
  {author} {\bibfnamefont {A.}~\bibnamefont {Beyer}}, \bibinfo {author}
  {\bibfnamefont {R.}~\bibnamefont {Holzwarth}}, \bibinfo {author}
  {\bibfnamefont {T.}~\bibnamefont {Udem}}, \bibinfo {author} {\bibfnamefont
  {T.}~\bibnamefont {Wilken}}, \bibinfo {author} {\bibfnamefont
  {N.}~\bibnamefont {Kolachevsky}}, \bibinfo {author} {\bibfnamefont
  {M.}~\bibnamefont {Abgrall}}, \bibinfo {author} {\bibfnamefont
  {D.}~\bibnamefont {Rovera}}, \bibinfo {author} {\bibfnamefont
  {C.}~\bibnamefont {Salomon}}, \bibinfo {author} {\bibfnamefont
  {P.}~\bibnamefont {Laurent}}, \bibinfo {author} {\bibfnamefont
  {G.}~\bibnamefont {Grosche}}, \bibinfo {author} {\bibfnamefont
  {O.}~\bibnamefont {Terra}}, \bibinfo {author} {\bibfnamefont
  {T.}~\bibnamefont {Legero}}, \bibinfo {author} {\bibfnamefont
  {H.}~\bibnamefont {Schnatz}}, \bibinfo {author} {\bibfnamefont
  {S.}~\bibnamefont {Weyers}}, \bibinfo {author} {\bibfnamefont
  {B.}~\bibnamefont {Altschul}},\ and\ \bibinfo {author} {\bibfnamefont
  {T.~W.}\ \bibnamefont {H\"ansch}},\ }\bibfield  {title} {\bibinfo {title}
  {Precision measurement of the hydrogen $1s\mathrm{\text{\ensuremath{-}}}2s$
  frequency via a 920-km fiber link},\ }\href
  {https://doi.org/10.1103/PhysRevLett.110.230801} {\bibfield  {journal}
  {\bibinfo  {journal} {Phys. Rev. Lett.}\ }\textbf {\bibinfo {volume} {110}},\
  \bibinfo {pages} {230801} (\bibinfo {year} {2013})}\BibitemShut {NoStop}%
\bibitem [{\citenamefont {Pruttivarasin}\ \emph {et~al.}(2015)\citenamefont
  {Pruttivarasin}, \citenamefont {Ramm}, \citenamefont {Porsev}, \citenamefont
  {Tupitsyn}, \citenamefont {Safronova}, \citenamefont {Hohensee},\ and\
  \citenamefont {H{\"a}ffner}}]{pruttivarasin15}%
  \BibitemOpen
  \bibfield  {author} {\bibinfo {author} {\bibfnamefont {T.}~\bibnamefont
  {Pruttivarasin}}, \bibinfo {author} {\bibfnamefont {M.}~\bibnamefont {Ramm}},
  \bibinfo {author} {\bibfnamefont {S.}~\bibnamefont {Porsev}}, \bibinfo
  {author} {\bibfnamefont {I.}~\bibnamefont {Tupitsyn}}, \bibinfo {author}
  {\bibfnamefont {M.}~\bibnamefont {Safronova}}, \bibinfo {author}
  {\bibfnamefont {M.}~\bibnamefont {Hohensee}},\ and\ \bibinfo {author}
  {\bibfnamefont {H.}~\bibnamefont {H{\"a}ffner}},\ }\bibfield  {title}
  {\bibinfo {title} {{Michelson-Morley} analogue for electrons using trapped
  ions to test {Lorentz} symmetry},\ }\href@noop {} {\bibfield  {journal}
  {\bibinfo  {journal} {Nature}\ }\textbf {\bibinfo {volume} {517}},\ \bibinfo
  {pages} {592} (\bibinfo {year} {2015})}\BibitemShut {NoStop}%
\bibitem [{\citenamefont {Delva}\ \emph {et~al.}(2018)\citenamefont {Delva},
  \citenamefont {Puchades}, \citenamefont {Sch\"onemann}, \citenamefont
  {Dilssner}, \citenamefont {Courde}, \citenamefont {Bertone}, \citenamefont
  {Gonzalez}, \citenamefont {Hees}, \citenamefont {Le~Poncin-Lafitte},
  \citenamefont {Meynadier}, \citenamefont {Prieto-Cerdeira}, \citenamefont
  {Sohet}, \citenamefont {Ventura-Traveset},\ and\ \citenamefont
  {Wolf}}]{delva18}%
  \BibitemOpen
  \bibfield  {author} {\bibinfo {author} {\bibfnamefont {P.}~\bibnamefont
  {Delva}}, \bibinfo {author} {\bibfnamefont {N.}~\bibnamefont {Puchades}},
  \bibinfo {author} {\bibfnamefont {E.}~\bibnamefont {Sch\"onemann}}, \bibinfo
  {author} {\bibfnamefont {F.}~\bibnamefont {Dilssner}}, \bibinfo {author}
  {\bibfnamefont {C.}~\bibnamefont {Courde}}, \bibinfo {author} {\bibfnamefont
  {S.}~\bibnamefont {Bertone}}, \bibinfo {author} {\bibfnamefont
  {F.}~\bibnamefont {Gonzalez}}, \bibinfo {author} {\bibfnamefont
  {A.}~\bibnamefont {Hees}}, \bibinfo {author} {\bibfnamefont {C.}~\bibnamefont
  {Le~Poncin-Lafitte}}, \bibinfo {author} {\bibfnamefont {F.}~\bibnamefont
  {Meynadier}}, \bibinfo {author} {\bibfnamefont {R.}~\bibnamefont
  {Prieto-Cerdeira}}, \bibinfo {author} {\bibfnamefont {B.}~\bibnamefont
  {Sohet}}, \bibinfo {author} {\bibfnamefont {J.}~\bibnamefont
  {Ventura-Traveset}},\ and\ \bibinfo {author} {\bibfnamefont {P.}~\bibnamefont
  {Wolf}},\ }\bibfield  {title} {\bibinfo {title} {Gravitational redshift test
  using eccentric {Galileo} satellites},\ }\href
  {https://doi.org/10.1103/PhysRevLett.121.231101} {\bibfield  {journal}
  {\bibinfo  {journal} {Phys. Rev. Lett.}\ }\textbf {\bibinfo {volume} {121}},\
  \bibinfo {pages} {231101} (\bibinfo {year} {2018})}\BibitemShut {NoStop}%
\bibitem [{\citenamefont {Herrmann}\ \emph {et~al.}(2018)\citenamefont
  {Herrmann}, \citenamefont {Finke}, \citenamefont {L\"ulf}, \citenamefont
  {Kichakova}, \citenamefont {Puetzfeld}, \citenamefont {Knickmann},
  \citenamefont {List}, \citenamefont {Rievers}, \citenamefont {Giorgi},
  \citenamefont {G\"unther}, \citenamefont {Dittus}, \citenamefont
  {Prieto-Cerdeira}, \citenamefont {Dilssner}, \citenamefont {Gonzalez},
  \citenamefont {Sch\"onemann}, \citenamefont {Ventura-Traveset},\ and\
  \citenamefont {L\"ammerzahl}}]{hermann18}%
  \BibitemOpen
  \bibfield  {author} {\bibinfo {author} {\bibfnamefont {S.}~\bibnamefont
  {Herrmann}}, \bibinfo {author} {\bibfnamefont {F.}~\bibnamefont {Finke}},
  \bibinfo {author} {\bibfnamefont {M.}~\bibnamefont {L\"ulf}}, \bibinfo
  {author} {\bibfnamefont {O.}~\bibnamefont {Kichakova}}, \bibinfo {author}
  {\bibfnamefont {D.}~\bibnamefont {Puetzfeld}}, \bibinfo {author}
  {\bibfnamefont {D.}~\bibnamefont {Knickmann}}, \bibinfo {author}
  {\bibfnamefont {M.}~\bibnamefont {List}}, \bibinfo {author} {\bibfnamefont
  {B.}~\bibnamefont {Rievers}}, \bibinfo {author} {\bibfnamefont
  {G.}~\bibnamefont {Giorgi}}, \bibinfo {author} {\bibfnamefont
  {C.}~\bibnamefont {G\"unther}}, \bibinfo {author} {\bibfnamefont
  {H.}~\bibnamefont {Dittus}}, \bibinfo {author} {\bibfnamefont
  {R.}~\bibnamefont {Prieto-Cerdeira}}, \bibinfo {author} {\bibfnamefont
  {F.}~\bibnamefont {Dilssner}}, \bibinfo {author} {\bibfnamefont
  {F.}~\bibnamefont {Gonzalez}}, \bibinfo {author} {\bibfnamefont
  {E.}~\bibnamefont {Sch\"onemann}}, \bibinfo {author} {\bibfnamefont
  {J.}~\bibnamefont {Ventura-Traveset}},\ and\ \bibinfo {author} {\bibfnamefont
  {C.}~\bibnamefont {L\"ammerzahl}},\ }\bibfield  {title} {\bibinfo {title}
  {Test of the gravitational redshift with {Galileo} satellites in an eccentric
  orbit},\ }\href {https://doi.org/10.1103/PhysRevLett.121.231102} {\bibfield
  {journal} {\bibinfo  {journal} {Phys. Rev. Lett.}\ }\textbf {\bibinfo
  {volume} {121}},\ \bibinfo {pages} {231102} (\bibinfo {year}
  {2018})}\BibitemShut {NoStop}%
\bibitem [{\citenamefont {Chou}\ \emph {et~al.}(2010)\citenamefont {Chou},
  \citenamefont {Hume}, \citenamefont {Rosenband},\ and\ \citenamefont
  {Wineland}}]{chou10}%
  \BibitemOpen
  \bibfield  {author} {\bibinfo {author} {\bibfnamefont {C.-W.}\ \bibnamefont
  {Chou}}, \bibinfo {author} {\bibfnamefont {D.}~\bibnamefont {Hume}}, \bibinfo
  {author} {\bibfnamefont {T.}~\bibnamefont {Rosenband}},\ and\ \bibinfo
  {author} {\bibfnamefont {D.}~\bibnamefont {Wineland}},\ }\bibfield  {title}
  {\bibinfo {title} {Optical clocks and relativity},\ }\href@noop {} {\bibfield
   {journal} {\bibinfo  {journal} {Science}\ }\textbf {\bibinfo {volume}
  {329}},\ \bibinfo {pages} {1630} (\bibinfo {year} {2010})}\BibitemShut
  {NoStop}%
\bibitem [{\citenamefont {Takamoto}\ \emph {et~al.}(2020)\citenamefont
  {Takamoto}, \citenamefont {Ushijima}, \citenamefont {Ohmae}, \citenamefont
  {Yahagi}, \citenamefont {Kokado}, \citenamefont {Shinkai},\ and\
  \citenamefont {Katori}}]{takamoto20}%
  \BibitemOpen
  \bibfield  {author} {\bibinfo {author} {\bibfnamefont {M.}~\bibnamefont
  {Takamoto}}, \bibinfo {author} {\bibfnamefont {I.}~\bibnamefont {Ushijima}},
  \bibinfo {author} {\bibfnamefont {N.}~\bibnamefont {Ohmae}}, \bibinfo
  {author} {\bibfnamefont {T.}~\bibnamefont {Yahagi}}, \bibinfo {author}
  {\bibfnamefont {K.}~\bibnamefont {Kokado}}, \bibinfo {author} {\bibfnamefont
  {H.}~\bibnamefont {Shinkai}},\ and\ \bibinfo {author} {\bibfnamefont
  {H.}~\bibnamefont {Katori}},\ }\bibfield  {title} {\bibinfo {title} {Test of
  general relativity by a pair of transportable optical lattice clocks},\
  }\href {https://doi.org/10.1038/s41566-020-0619-8} {\bibfield  {journal}
  {\bibinfo  {journal} {Nature Phot.}\ }\textbf {\bibinfo {volume} {14}},\
  \bibinfo {pages} {411} (\bibinfo {year} {2020})}\BibitemShut {NoStop}%
\bibitem [{\citenamefont {M{\"u}ller}\ \emph {et~al.}(2010)\citenamefont
  {M{\"u}ller}, \citenamefont {Peters},\ and\ \citenamefont {Chu}}]{mueller10}%
  \BibitemOpen
  \bibfield  {author} {\bibinfo {author} {\bibfnamefont {H.}~\bibnamefont
  {M{\"u}ller}}, \bibinfo {author} {\bibfnamefont {A.}~\bibnamefont {Peters}},\
  and\ \bibinfo {author} {\bibfnamefont {S.}~\bibnamefont {Chu}},\ }\bibfield
  {title} {\bibinfo {title} {A precision measurement of the gravitational
  redshift by the interference of matter waves},\ }\href@noop {} {\bibfield
  {journal} {\bibinfo  {journal} {Nature}\ }\textbf {\bibinfo {volume} {463}},\
  \bibinfo {pages} {926} (\bibinfo {year} {2010})}\BibitemShut {NoStop}%
\bibitem [{\citenamefont {Wolf}\ \emph {et~al.}(2011)\citenamefont {Wolf},
  \citenamefont {Blanchet}, \citenamefont {Bord{\'e}}, \citenamefont {Reynaud},
  \citenamefont {Salomon},\ and\ \citenamefont {Cohen-Tannoudji}}]{wolf11a}%
  \BibitemOpen
  \bibfield  {author} {\bibinfo {author} {\bibfnamefont {P.}~\bibnamefont
  {Wolf}}, \bibinfo {author} {\bibfnamefont {L.}~\bibnamefont {Blanchet}},
  \bibinfo {author} {\bibfnamefont {C.~J.}\ \bibnamefont {Bord{\'e}}}, \bibinfo
  {author} {\bibfnamefont {S.}~\bibnamefont {Reynaud}}, \bibinfo {author}
  {\bibfnamefont {C.}~\bibnamefont {Salomon}},\ and\ \bibinfo {author}
  {\bibfnamefont {C.}~\bibnamefont {Cohen-Tannoudji}},\ }\bibfield  {title}
  {\bibinfo {title} {Does an atom interferometer test the gravitational
  redshift at the {Compton} frequency?},\ }\href@noop {} {\bibfield  {journal}
  {\bibinfo  {journal} {Class. Quant. Grav.}\ }\textbf {\bibinfo {volume}
  {28}},\ \bibinfo {pages} {145017} (\bibinfo {year} {2011})}\BibitemShut
  {NoStop}%
\bibitem [{\citenamefont {Schleich}\ \emph {et~al.}(2013)\citenamefont
  {Schleich}, \citenamefont {Greenberger},\ and\ \citenamefont
  {Rasel}}]{schleich13a}%
  \BibitemOpen
  \bibfield  {author} {\bibinfo {author} {\bibfnamefont {W.~P.}\ \bibnamefont
  {Schleich}}, \bibinfo {author} {\bibfnamefont {D.~M.}\ \bibnamefont
  {Greenberger}},\ and\ \bibinfo {author} {\bibfnamefont {E.~M.}\ \bibnamefont
  {Rasel}},\ }\bibfield  {title} {\bibinfo {title} {A representation-free
  description of the {Kasevich-Chu} interferometer: a resolution of the
  redshift controversy},\ }\href
  {http://stacks.iop.org/1367-2630/15/i=1/a=013007} {\bibfield  {journal}
  {\bibinfo  {journal} {New J. Phys.}\ }\textbf {\bibinfo {volume} {15}},\
  \bibinfo {pages} {013007} (\bibinfo {year} {2013})}\BibitemShut {NoStop}%
\bibitem [{\citenamefont {Sinha}\ and\ \citenamefont {Samuel}(2011)}]{sinha11}%
  \BibitemOpen
  \bibfield  {author} {\bibinfo {author} {\bibfnamefont {S.}~\bibnamefont
  {Sinha}}\ and\ \bibinfo {author} {\bibfnamefont {J.}~\bibnamefont {Samuel}},\
  }\bibfield  {title} {\bibinfo {title} {Atom interferometry and the
  gravitational redshift},\ }\href
  {https://doi.org/10.1088/0264-9381/28/14/145018} {\bibfield  {journal}
  {\bibinfo  {journal} {Class. Quant. Grav.}\ }\textbf {\bibinfo {volume}
  {28}},\ \bibinfo {pages} {145018} (\bibinfo {year} {2011})}\BibitemShut
  {NoStop}%
\bibitem [{\citenamefont {Roura}(2020)}]{roura20a}%
  \BibitemOpen
  \bibfield  {author} {\bibinfo {author} {\bibfnamefont {A.}~\bibnamefont
  {Roura}},\ }\bibfield  {title} {\bibinfo {title} {Gravitational redshift in
  quantum-clock interferometry},\ }\href
  {https://doi.org/10.1103/PhysRevX.10.021014} {\bibfield  {journal} {\bibinfo
  {journal} {Phys. Rev. X}\ }\textbf {\bibinfo {volume} {10}},\ \bibinfo
  {pages} {021014} (\bibinfo {year} {2020})}\BibitemShut {NoStop}%
\bibitem [{\citenamefont {Loriani}\ \emph {et~al.}(2019)\citenamefont
  {Loriani}, \citenamefont {Friedrich}, \citenamefont {Ufrecht}, \citenamefont
  {Di~Pumpo}, \citenamefont {Kleinert}, \citenamefont {Abend}, \citenamefont
  {Gaaloul}, \citenamefont {Meiners}, \citenamefont {Schubert}, \citenamefont
  {Tell}, \citenamefont {Wodey}, \citenamefont {Zych}, \citenamefont {Ertmer},
  \citenamefont {Roura}, \citenamefont {Schlippert}, \citenamefont {Schleich},
  \citenamefont {Rasel},\ and\ \citenamefont {Giese}}]{loriani19}%
  \BibitemOpen
  \bibfield  {author} {\bibinfo {author} {\bibfnamefont {S.}~\bibnamefont
  {Loriani}}, \bibinfo {author} {\bibfnamefont {A.}~\bibnamefont {Friedrich}},
  \bibinfo {author} {\bibfnamefont {C.}~\bibnamefont {Ufrecht}}, \bibinfo
  {author} {\bibfnamefont {F.}~\bibnamefont {Di~Pumpo}}, \bibinfo {author}
  {\bibfnamefont {S.}~\bibnamefont {Kleinert}}, \bibinfo {author}
  {\bibfnamefont {S.}~\bibnamefont {Abend}}, \bibinfo {author} {\bibfnamefont
  {N.}~\bibnamefont {Gaaloul}}, \bibinfo {author} {\bibfnamefont
  {C.}~\bibnamefont {Meiners}}, \bibinfo {author} {\bibfnamefont
  {C.}~\bibnamefont {Schubert}}, \bibinfo {author} {\bibfnamefont
  {D.}~\bibnamefont {Tell}}, \bibinfo {author} {\bibfnamefont
  {{\'E}.}~\bibnamefont {Wodey}}, \bibinfo {author} {\bibfnamefont
  {M.}~\bibnamefont {Zych}}, \bibinfo {author} {\bibfnamefont {W.}~\bibnamefont
  {Ertmer}}, \bibinfo {author} {\bibfnamefont {A.}~\bibnamefont {Roura}},
  \bibinfo {author} {\bibfnamefont {D.}~\bibnamefont {Schlippert}}, \bibinfo
  {author} {\bibfnamefont {W.~P.}\ \bibnamefont {Schleich}}, \bibinfo {author}
  {\bibfnamefont {E.~M.}\ \bibnamefont {Rasel}},\ and\ \bibinfo {author}
  {\bibfnamefont {E.}~\bibnamefont {Giese}},\ }\bibfield  {title} {\bibinfo
  {title} {Interference of clocks: A quantum twin paradox},\ }\href
  {https://doi.org/10.1126/sciadv.aax8966} {\bibfield  {journal} {\bibinfo
  {journal} {Science Advances}\ }\textbf {\bibinfo {volume} {5}},\ \bibinfo
  {pages} {eaax8966} (\bibinfo {year} {2019})}\BibitemShut {NoStop}%
\bibitem [{\citenamefont {Ufrecht}\ \emph {et~al.}()\citenamefont {Ufrecht},
  \citenamefont {Di~Pumpo}, \citenamefont {Friedrich}, \citenamefont {Roura},
  \citenamefont {Schubert}, \citenamefont {Schlippert}, \citenamefont
  {Schleich}, \citenamefont {Rasel},\ and\ \citenamefont {Giese}}]{ufrecht20}%
  \BibitemOpen
  \bibfield  {author} {\bibinfo {author} {\bibfnamefont {C.}~\bibnamefont
  {Ufrecht}}, \bibinfo {author} {\bibfnamefont {F.}~\bibnamefont {Di~Pumpo}},
  \bibinfo {author} {\bibfnamefont {A.}~\bibnamefont {Friedrich}}, \bibinfo
  {author} {\bibfnamefont {A.}~\bibnamefont {Roura}}, \bibinfo {author}
  {\bibfnamefont {C.}~\bibnamefont {Schubert}}, \bibinfo {author}
  {\bibfnamefont {D.}~\bibnamefont {Schlippert}}, \bibinfo {author}
  {\bibfnamefont {W.~P.}\ \bibnamefont {Schleich}}, \bibinfo {author}
  {\bibfnamefont {E.~M.}\ \bibnamefont {Rasel}},\ and\ \bibinfo {author}
  {\bibfnamefont {E.}~\bibnamefont {Giese}},\ }\bibfield  {title} {\bibinfo
  {title} {An atom interferometer testing the universality of free fall and
  gravitational redshift},\ }\href {https://arxiv.org/abs/2001.09754} {\bibinfo
   {journal} {\texttt{arXiv:2001.09754}}\ }\BibitemShut {NoStop}%
\bibitem [{\citenamefont {Hartwig}\ \emph {et~al.}(2015)\citenamefont
  {Hartwig}, \citenamefont {Abend}, \citenamefont {Schubert}, \citenamefont
  {Schlippert}, \citenamefont {Ahlers}, \citenamefont {Posso-Trujillo},
  \citenamefont {Gaaloul}, \citenamefont {Ertmer},\ and\ \citenamefont
  {Rasel}}]{hartwig15}%
  \BibitemOpen
\bibfield  {journal} {  }\bibfield  {author} {\bibinfo {author} {\bibfnamefont
  {J.}~\bibnamefont {Hartwig}}, \bibinfo {author} {\bibfnamefont
  {S.}~\bibnamefont {Abend}}, \bibinfo {author} {\bibfnamefont
  {C.}~\bibnamefont {Schubert}}, \bibinfo {author} {\bibfnamefont
  {D.}~\bibnamefont {Schlippert}}, \bibinfo {author} {\bibfnamefont
  {H.}~\bibnamefont {Ahlers}}, \bibinfo {author} {\bibfnamefont
  {K.}~\bibnamefont {Posso-Trujillo}}, \bibinfo {author} {\bibfnamefont
  {N.}~\bibnamefont {Gaaloul}}, \bibinfo {author} {\bibfnamefont
  {W.}~\bibnamefont {Ertmer}},\ and\ \bibinfo {author} {\bibfnamefont {E.~M.}\
  \bibnamefont {Rasel}},\ }\bibfield  {title} {\bibinfo {title} {Testing the
  universality of free fall with rubidium and ytterbium in a very large
  baseline atom interferometer},\ }\href
  {https://doi.org/10.1088/1367-2630/17/3/035011} {\bibfield  {journal}
  {\bibinfo  {journal} {New J. Phys.}\ }\textbf {\bibinfo {volume} {17}},\
  \bibinfo {pages} {035011} (\bibinfo {year} {2015})}\BibitemShut {NoStop}%
\bibitem [{\citenamefont {Schlippert}\ \emph {et~al.}(2020)\citenamefont
  {Schlippert}, \citenamefont {Meiners}, \citenamefont {Rengelink},
  \citenamefont {Schubert}, \citenamefont {Tell}, \citenamefont {Wodey},
  \citenamefont {Zipfel}, \citenamefont {Ertmer},\ and\ \citenamefont
  {Rasel}}]{schlippert20}%
  \BibitemOpen
  \bibfield  {author} {\bibinfo {author} {\bibfnamefont {D.}~\bibnamefont
  {Schlippert}}, \bibinfo {author} {\bibfnamefont {C.}~\bibnamefont {Meiners}},
  \bibinfo {author} {\bibfnamefont {R.}~\bibnamefont {Rengelink}}, \bibinfo
  {author} {\bibfnamefont {C.}~\bibnamefont {Schubert}}, \bibinfo {author}
  {\bibfnamefont {D.}~\bibnamefont {Tell}}, \bibinfo {author} {\bibfnamefont
  {E.}~\bibnamefont {Wodey}}, \bibinfo {author} {\bibfnamefont
  {K.}~\bibnamefont {Zipfel}}, \bibinfo {author} {\bibfnamefont
  {W.}~\bibnamefont {Ertmer}},\ and\ \bibinfo {author} {\bibfnamefont
  {E.}~\bibnamefont {Rasel}},\ }\bibfield  {title} {\bibinfo {title}
  {Matter-wave interferometry for inertial sensing and tests of fundamental
  physics},\ }in\ \href {https://doi.org/10.1142/9789811213984_0010} {\emph
  {\bibinfo {booktitle} {Proceedings of the Eighth Meeting on CPT and Lorentz
  Symmetry}}}\ (\bibinfo  {publisher} {World Scientific},\ \bibinfo {year}
  {2020})\BibitemShut {NoStop}%
\bibitem [{\citenamefont {Bord\'{e}}(1992)}]{borde92}%
  \BibitemOpen
  \bibfield  {author} {\bibinfo {author} {\bibfnamefont {C.~J.}\ \bibnamefont
  {Bord\'{e}}},\ }\bibfield  {title} {\bibinfo {title} {Propagation of laser
  beams and of atomic systems},\ }in\ \href@noop {} {\emph {\bibinfo
  {booktitle} {Les Houches lectures on Fundamental Systems in Quantum
  Optics}}},\ \bibinfo {editor} {edited by\ \bibinfo {editor} {\bibfnamefont
  {J.}~\bibnamefont {Dalibard}}}\ (\bibinfo  {publisher} {Elsevier},\ \bibinfo
  {year} {1992})\BibitemShut {NoStop}%
\bibitem [{\citenamefont {Antoine}\ and\ \citenamefont
  {Bord\'{e}}(2003)}]{antoine03b}%
  \BibitemOpen
  \bibfield  {author} {\bibinfo {author} {\bibfnamefont {C.}~\bibnamefont
  {Antoine}}\ and\ \bibinfo {author} {\bibfnamefont {C.~J.}\ \bibnamefont
  {Bord\'{e}}},\ }\bibfield  {title} {\bibinfo {title} {Quantum theory of
  atomic clocks and gravito-inertial sensors: an update},\ }\href
  {http://stacks.iop.org/1464-4266/5/i=2/a=380} {\bibfield  {journal} {\bibinfo
   {journal} {J. Opt. B: Quantum and Semiclass. Opt.}\ }\textbf {\bibinfo
  {volume} {5}},\ \bibinfo {pages} {S199} (\bibinfo {year} {2003})}\BibitemShut
  {NoStop}%
\bibitem [{\citenamefont {Hogan}\ \emph {et~al.}()\citenamefont {Hogan},
  \citenamefont {Johnson},\ and\ \citenamefont {Kasevich}}]{hogan08}%
  \BibitemOpen
  \bibfield  {author} {\bibinfo {author} {\bibfnamefont {J.~M.}\ \bibnamefont
  {Hogan}}, \bibinfo {author} {\bibfnamefont {D.~M.~S.}\ \bibnamefont
  {Johnson}},\ and\ \bibinfo {author} {\bibfnamefont {M.~A.}\ \bibnamefont
  {Kasevich}},\ }\bibfield  {title} {\bibinfo {title} {Light-pulse atom
  interferometry},\ }\href {https://arxiv.org/abs/0806.3261} {\bibinfo
  {journal} {\texttt{arXiv:0806.3261}}\ }\BibitemShut {NoStop}%
\bibitem [{\citenamefont {Roura}\ \emph {et~al.}(2014)\citenamefont {Roura},
  \citenamefont {Zeller},\ and\ \citenamefont {Schleich}}]{roura14}%
  \BibitemOpen
\bibfield  {journal} {  }\bibfield  {author} {\bibinfo {author} {\bibfnamefont
  {A.}~\bibnamefont {Roura}}, \bibinfo {author} {\bibfnamefont
  {W.}~\bibnamefont {Zeller}},\ and\ \bibinfo {author} {\bibfnamefont {W.~P.}\
  \bibnamefont {Schleich}},\ }\bibfield  {title} {\bibinfo {title} {Overcoming
  loss of contrast in atom interferometry due to gravity gradients},\ }\href
  {http://stacks.iop.org/1367-2630/16/i=12/a=123012} {\bibfield  {journal}
  {\bibinfo  {journal} {New J. Phys.}\ }\textbf {\bibinfo {volume} {16}},\
  \bibinfo {pages} {123012} (\bibinfo {year} {2014})}\BibitemShut {NoStop}%
\bibitem [{\citenamefont {Misner}\ \emph {et~al.}(1973)\citenamefont {Misner},
  \citenamefont {Thorne},\ and\ \citenamefont {Wheeler}}]{misner73}%
  \BibitemOpen
  \bibfield  {author} {\bibinfo {author} {\bibfnamefont {C.~W.}\ \bibnamefont
  {Misner}}, \bibinfo {author} {\bibfnamefont {K.~S.}\ \bibnamefont {Thorne}},\
  and\ \bibinfo {author} {\bibfnamefont {J.~A.}\ \bibnamefont {Wheeler}},\
  }\href@noop {} {\emph {\bibinfo {title} {Gravitation}}}\ (\bibinfo
  {publisher} {Freeman},\ \bibinfo {address} {San Francisco},\ \bibinfo {year}
  {1973})\BibitemShut {NoStop}%
\bibitem [{\citenamefont {Zych}\ \emph {et~al.}(2011)\citenamefont {Zych},
  \citenamefont {Costa}, \citenamefont {Pikovski},\ and\ \citenamefont
  {Brukner}}]{zych11}%
  \BibitemOpen
  \bibfield  {author} {\bibinfo {author} {\bibfnamefont {M.}~\bibnamefont
  {Zych}}, \bibinfo {author} {\bibfnamefont {F.}~\bibnamefont {Costa}},
  \bibinfo {author} {\bibfnamefont {I.}~\bibnamefont {Pikovski}},\ and\
  \bibinfo {author} {\bibfnamefont {{\v{C}}.}~\bibnamefont {Brukner}},\
  }\bibfield  {title} {\bibinfo {title} {Quantum interferometric visibility as
  a witness of general relativistic proper time},\ }\href@noop {} {\bibfield
  {journal} {\bibinfo  {journal} {Nat. Comm.}\ }\textbf {\bibinfo {volume}
  {2}},\ \bibinfo {pages} {505} (\bibinfo {year} {2011})}\BibitemShut {NoStop}%
\bibitem [{\citenamefont {Bord\'{e}}(1989)}]{borde89}%
  \BibitemOpen
  \bibfield  {author} {\bibinfo {author} {\bibfnamefont {C.~J.}\ \bibnamefont
  {Bord\'{e}}},\ }\bibfield  {title} {\bibinfo {title} {Atomic interferometry
  with internal state labelling},\ }\href
  {https://doi.org/10.1016/0375-9601(89)90537-9} {\bibfield  {journal}
  {\bibinfo  {journal} {Phys. Lett. A}\ }\textbf {\bibinfo {volume} {140}},\
  \bibinfo {pages} {10 } (\bibinfo {year} {1989})}\BibitemShut {NoStop}%
\bibitem [{\citenamefont {Kozuma}\ \emph {et~al.}(1999)\citenamefont {Kozuma},
  \citenamefont {Deng}, \citenamefont {Hagley}, \citenamefont {Wen},
  \citenamefont {Lutwak}, \citenamefont {Helmerson}, \citenamefont {Rolston},\
  and\ \citenamefont {Phillips}}]{kozuma99}%
  \BibitemOpen
  \bibfield  {author} {\bibinfo {author} {\bibfnamefont {M.}~\bibnamefont
  {Kozuma}}, \bibinfo {author} {\bibfnamefont {L.}~\bibnamefont {Deng}},
  \bibinfo {author} {\bibfnamefont {E.~W.}\ \bibnamefont {Hagley}}, \bibinfo
  {author} {\bibfnamefont {J.}~\bibnamefont {Wen}}, \bibinfo {author}
  {\bibfnamefont {R.}~\bibnamefont {Lutwak}}, \bibinfo {author} {\bibfnamefont
  {K.}~\bibnamefont {Helmerson}}, \bibinfo {author} {\bibfnamefont {S.~L.}\
  \bibnamefont {Rolston}},\ and\ \bibinfo {author} {\bibfnamefont {W.~D.}\
  \bibnamefont {Phillips}},\ }\bibfield  {title} {\bibinfo {title} {Coherent
  splitting of {Bose-Einstein} condensed atoms with optically induced {Bragg}
  diffraction},\ }\href {https://doi.org/10.1103/PhysRevLett.82.871} {\bibfield
   {journal} {\bibinfo  {journal} {Phys. Rev. Lett.}\ }\textbf {\bibinfo
  {volume} {82}},\ \bibinfo {pages} {871} (\bibinfo {year} {1999})}\BibitemShut
  {NoStop}%
\bibitem [{\citenamefont {Ye}\ \emph {et~al.}(2008)\citenamefont {Ye},
  \citenamefont {Kimble},\ and\ \citenamefont {Katori}}]{ye08}%
  \BibitemOpen
  \bibfield  {author} {\bibinfo {author} {\bibfnamefont {J.}~\bibnamefont
  {Ye}}, \bibinfo {author} {\bibfnamefont {H.~J.}\ \bibnamefont {Kimble}},\
  and\ \bibinfo {author} {\bibfnamefont {H.}~\bibnamefont {Katori}},\
  }\bibfield  {title} {\bibinfo {title} {Quantum state engineering and
  precision metrology using state-insensitive light traps},\ }\href@noop {}
  {\bibfield  {journal} {\bibinfo  {journal} {Science}\ }\textbf {\bibinfo
  {volume} {320}},\ \bibinfo {pages} {1734} (\bibinfo {year}
  {2008})}\BibitemShut {NoStop}%
\bibitem [{\citenamefont {Szigeti}\ \emph {et~al.}(2012)\citenamefont
  {Szigeti}, \citenamefont {Debs}, \citenamefont {Hope}, \citenamefont
  {Robins},\ and\ \citenamefont {Close}}]{szigeti12}%
  \BibitemOpen
  \bibfield  {author} {\bibinfo {author} {\bibfnamefont {S.~S.}\ \bibnamefont
  {Szigeti}}, \bibinfo {author} {\bibfnamefont {J.~E.}\ \bibnamefont {Debs}},
  \bibinfo {author} {\bibfnamefont {J.~J.}\ \bibnamefont {Hope}}, \bibinfo
  {author} {\bibfnamefont {N.~P.}\ \bibnamefont {Robins}},\ and\ \bibinfo
  {author} {\bibfnamefont {J.~D.}\ \bibnamefont {Close}},\ }\bibfield  {title}
  {\bibinfo {title} {Why momentum width matters for atom interferometry with
  {Bragg} pulses},\ }\href {https://doi.org/10.1088/1367-2630/14/2/023009}
  {\bibfield  {journal} {\bibinfo  {journal} {New J. Phys.}\ }\textbf {\bibinfo
  {volume} {14}},\ \bibinfo {pages} {023009} (\bibinfo {year}
  {2012})}\BibitemShut {NoStop}%
\bibitem [{\citenamefont {Abend}\ \emph {et~al.}(2020)\citenamefont {Abend},
  \citenamefont {Gersemann}, \citenamefont {Schubert}, \citenamefont
  {Schlippert}, \citenamefont {Rasel}, \citenamefont {Zimmermann},
  \citenamefont {Efremov}, \citenamefont {Roura}, \citenamefont {Narducci},\
  and\ \citenamefont {Schleich}}]{abend20}%
  \BibitemOpen
  \bibfield  {author} {\bibinfo {author} {\bibfnamefont {S.}~\bibnamefont
  {Abend}}, \bibinfo {author} {\bibfnamefont {M.}~\bibnamefont {Gersemann}},
  \bibinfo {author} {\bibfnamefont {C.}~\bibnamefont {Schubert}}, \bibinfo
  {author} {\bibfnamefont {D.}~\bibnamefont {Schlippert}}, \bibinfo {author}
  {\bibfnamefont {E.~M.}\ \bibnamefont {Rasel}}, \bibinfo {author}
  {\bibfnamefont {M.}~\bibnamefont {Zimmermann}}, \bibinfo {author}
  {\bibfnamefont {M.~A.}\ \bibnamefont {Efremov}}, \bibinfo {author}
  {\bibfnamefont {A.}~\bibnamefont {Roura}}, \bibinfo {author} {\bibfnamefont
  {F.~A.}\ \bibnamefont {Narducci}},\ and\ \bibinfo {author} {\bibfnamefont
  {W.~P.}\ \bibnamefont {Schleich}},\ }\bibfield  {title} {\bibinfo {title}
  {Atom interferometry and its applications},\ }in\ \href
  {https://arxiv.org/abs/2001.10976} {\emph {\bibinfo {booktitle} {{Proceedings
  of the International School of Physics ``Enrico Fermi'' Course 197
  ``Foundations of Quantum Theory''}}}},\ \bibinfo {editor} {edited by\
  \bibinfo {editor} {\bibfnamefont {E.~M.}\ \bibnamefont {Rasel}}, \bibinfo
  {editor} {\bibfnamefont {W.~P.}\ \bibnamefont {Schleich}},\ and\ \bibinfo
  {editor} {\bibfnamefont {S.}~\bibnamefont {W\"olk}}}\ (\bibinfo  {publisher}
  {IOS, Amsterdam; SIF, Bologna},\ \bibinfo {year} {2020})\ pp.\ \bibinfo
  {pages} {345--392}\BibitemShut {NoStop}%
\bibitem [{\citenamefont {Hu}\ \emph {et~al.}(2017)\citenamefont {Hu},
  \citenamefont {Poli}, \citenamefont {Salvi},\ and\ \citenamefont
  {Tino}}]{hu17}%
  \BibitemOpen
  \bibfield  {author} {\bibinfo {author} {\bibfnamefont {L.}~\bibnamefont
  {Hu}}, \bibinfo {author} {\bibfnamefont {N.}~\bibnamefont {Poli}}, \bibinfo
  {author} {\bibfnamefont {L.}~\bibnamefont {Salvi}},\ and\ \bibinfo {author}
  {\bibfnamefont {G.~M.}\ \bibnamefont {Tino}},\ }\bibfield  {title} {\bibinfo
  {title} {Atom interferometry with the {Sr} optical clock transition},\
  }\href@noop {} {\bibfield  {journal} {\bibinfo  {journal} {Phys. Rev. Lett.}\
  }\textbf {\bibinfo {volume} {119}},\ \bibinfo {pages} {263601} (\bibinfo
  {year} {2017})}\BibitemShut {NoStop}%
\bibitem [{\citenamefont {Hu}\ \emph {et~al.}(2019)\citenamefont {Hu},
  \citenamefont {Wang}, \citenamefont {Salvi}, \citenamefont {Tinsley},
  \citenamefont {Tino},\ and\ \citenamefont {Poli}}]{hu20}%
  \BibitemOpen
  \bibfield  {author} {\bibinfo {author} {\bibfnamefont {L.}~\bibnamefont
  {Hu}}, \bibinfo {author} {\bibfnamefont {E.}~\bibnamefont {Wang}}, \bibinfo
  {author} {\bibfnamefont {L.}~\bibnamefont {Salvi}}, \bibinfo {author}
  {\bibfnamefont {J.~N.}\ \bibnamefont {Tinsley}}, \bibinfo {author}
  {\bibfnamefont {G.~M.}\ \bibnamefont {Tino}},\ and\ \bibinfo {author}
  {\bibfnamefont {N.}~\bibnamefont {Poli}},\ }\bibfield  {title} {\bibinfo
  {title} {Sr atom interferometry with the optical clock transition as a
  gravimeter and a gravity gradiometer},\ }\href
  {https://doi.org/10.1088/1361-6382/ab4d18} {\bibfield  {journal} {\bibinfo
  {journal} {Class. Quant. Grav.}\ }\textbf {\bibinfo {volume} {37}},\ \bibinfo
  {pages} {014001} (\bibinfo {year} {2019})}\BibitemShut {NoStop}%
\bibitem [{\citenamefont {{MAGIS-100 at Fermilab}}()}]{magis}%
  \BibitemOpen
  \bibfield  {author} {\bibinfo {author} {\bibnamefont {{MAGIS-100 at
  Fermilab}}},\ }\href {http://qis.fnal.gov/magis-100} {\bibinfo  {journal}
  {http://qis.fnal.gov/magis-100}\ }\BibitemShut {NoStop}%
\bibitem [{\citenamefont {Badurina}\ \emph {et~al.}(2020)\citenamefont
  {Badurina}, \citenamefont {Bentine}, \citenamefont {Blas}, \citenamefont
  {Bongs}, \citenamefont {Bortoletto}, \citenamefont {Bowcock}, \citenamefont
  {Bridges}, \citenamefont {Bowden}, \citenamefont {Buchmueller}, \citenamefont
  {Burrage}, \citenamefont {Coleman}, \citenamefont {Elertas}, \citenamefont
  {Ellis}, \citenamefont {Foot}, \citenamefont {Gibson}, \citenamefont
  {Haehnelt}, \citenamefont {Harte}, \citenamefont {Hedges}, \citenamefont
  {Hobson}, \citenamefont {Holynski}, \citenamefont {Jones}, \citenamefont
  {Langlois}, \citenamefont {Lellouch}, \citenamefont {Lewicki}, \citenamefont
  {Maiolino}, \citenamefont {Majewski}, \citenamefont {Malik}, \citenamefont
  {March-Russell}, \citenamefont {McCabe}, \citenamefont {Newbold},
  \citenamefont {Sauer}, \citenamefont {Schneider}, \citenamefont {Shipsey},
  \citenamefont {Singh}, \citenamefont {Uchida}, \citenamefont {Valenzuela},
  \citenamefont {van~der Grinten}, \citenamefont {Vaskonen}, \citenamefont
  {Vossebeld}, \citenamefont {Weatherill},\ and\ \citenamefont
  {Wilmut}}]{badurina20}%
  \BibitemOpen
\bibfield  {journal} {  }\bibfield  {author} {\bibinfo {author} {\bibfnamefont
  {L.}~\bibnamefont {Badurina}}, \bibinfo {author} {\bibfnamefont
  {E.}~\bibnamefont {Bentine}}, \bibinfo {author} {\bibfnamefont
  {D.}~\bibnamefont {Blas}}, \bibinfo {author} {\bibfnamefont {K.}~\bibnamefont
  {Bongs}}, \bibinfo {author} {\bibfnamefont {D.}~\bibnamefont {Bortoletto}},
  \bibinfo {author} {\bibfnamefont {T.}~\bibnamefont {Bowcock}}, \bibinfo
  {author} {\bibfnamefont {K.}~\bibnamefont {Bridges}}, \bibinfo {author}
  {\bibfnamefont {W.}~\bibnamefont {Bowden}}, \bibinfo {author} {\bibfnamefont
  {O.}~\bibnamefont {Buchmueller}}, \bibinfo {author} {\bibfnamefont
  {C.}~\bibnamefont {Burrage}}, \bibinfo {author} {\bibfnamefont
  {J.}~\bibnamefont {Coleman}}, \bibinfo {author} {\bibfnamefont
  {G.}~\bibnamefont {Elertas}}, \bibinfo {author} {\bibfnamefont
  {J.}~\bibnamefont {Ellis}}, \bibinfo {author} {\bibfnamefont
  {C.}~\bibnamefont {Foot}}, \bibinfo {author} {\bibfnamefont {V.}~\bibnamefont
  {Gibson}}, \bibinfo {author} {\bibfnamefont {M.}~\bibnamefont {Haehnelt}},
  \bibinfo {author} {\bibfnamefont {T.}~\bibnamefont {Harte}}, \bibinfo
  {author} {\bibfnamefont {S.}~\bibnamefont {Hedges}}, \bibinfo {author}
  {\bibfnamefont {R.}~\bibnamefont {Hobson}}, \bibinfo {author} {\bibfnamefont
  {M.}~\bibnamefont {Holynski}}, \bibinfo {author} {\bibfnamefont
  {T.}~\bibnamefont {Jones}}, \bibinfo {author} {\bibfnamefont
  {M.}~\bibnamefont {Langlois}}, \bibinfo {author} {\bibfnamefont
  {S.}~\bibnamefont {Lellouch}}, \bibinfo {author} {\bibfnamefont
  {M.}~\bibnamefont {Lewicki}}, \bibinfo {author} {\bibfnamefont
  {R.}~\bibnamefont {Maiolino}}, \bibinfo {author} {\bibfnamefont
  {P.}~\bibnamefont {Majewski}}, \bibinfo {author} {\bibfnamefont
  {S.}~\bibnamefont {Malik}}, \bibinfo {author} {\bibfnamefont
  {J.}~\bibnamefont {March-Russell}}, \bibinfo {author} {\bibfnamefont
  {C.}~\bibnamefont {McCabe}}, \bibinfo {author} {\bibfnamefont
  {D.}~\bibnamefont {Newbold}}, \bibinfo {author} {\bibfnamefont
  {B.}~\bibnamefont {Sauer}}, \bibinfo {author} {\bibfnamefont
  {U.}~\bibnamefont {Schneider}}, \bibinfo {author} {\bibfnamefont
  {I.}~\bibnamefont {Shipsey}}, \bibinfo {author} {\bibfnamefont
  {Y.}~\bibnamefont {Singh}}, \bibinfo {author} {\bibfnamefont
  {M.}~\bibnamefont {Uchida}}, \bibinfo {author} {\bibfnamefont
  {T.}~\bibnamefont {Valenzuela}}, \bibinfo {author} {\bibfnamefont
  {M.}~\bibnamefont {van~der Grinten}}, \bibinfo {author} {\bibfnamefont
  {V.}~\bibnamefont {Vaskonen}}, \bibinfo {author} {\bibfnamefont
  {J.}~\bibnamefont {Vossebeld}}, \bibinfo {author} {\bibfnamefont
  {D.}~\bibnamefont {Weatherill}},\ and\ \bibinfo {author} {\bibfnamefont
  {I.}~\bibnamefont {Wilmut}},\ }\bibfield  {title} {\bibinfo {title} {{AION}:
  an atom interferometer observatory and network},\ }\href
  {https://doi.org/10.1088/1475-7516/2020/05/011} {\bibfield  {journal}
  {\bibinfo  {journal} {J. Cosmol. Astropart. Phys.}\ }\textbf {\bibinfo
  {volume} {2020}}\bibinfo  {number} { (05)},\ \bibinfo {pages}
  {011}}\BibitemShut {NoStop}%
\bibitem [{\citenamefont {del Aguila}\ \emph {et~al.}(2018)\citenamefont {del
  Aguila}, \citenamefont {Mazzoni}, \citenamefont {Hu}, \citenamefont {Salvi},
  \citenamefont {Tino},\ and\ \citenamefont {Poli}}]{del_aguila18}%
  \BibitemOpen
\bibfield  {number} {  }\bibfield  {author} {\bibinfo {author} {\bibfnamefont
  {R.~P.}\ \bibnamefont {del Aguila}}, \bibinfo {author} {\bibfnamefont
  {T.}~\bibnamefont {Mazzoni}}, \bibinfo {author} {\bibfnamefont
  {L.}~\bibnamefont {Hu}}, \bibinfo {author} {\bibfnamefont {L.}~\bibnamefont
  {Salvi}}, \bibinfo {author} {\bibfnamefont {G.~M.}\ \bibnamefont {Tino}},\
  and\ \bibinfo {author} {\bibfnamefont {N.}~\bibnamefont {Poli}},\ }\bibfield
  {title} {\bibinfo {title} {Bragg gravity-gradiometer using the
  {$^1S_0-{}^3P_1$} intercombination transition of $^{88}${Sr}},\ }\href@noop
  {} {\bibfield  {journal} {\bibinfo  {journal} {New J. Phys.}\ }\textbf
  {\bibinfo {volume} {20}},\ \bibinfo {pages} {043002} (\bibinfo {year}
  {2018})}\BibitemShut {NoStop}%
\bibitem [{\citenamefont {Plotkin-Swing}\ \emph {et~al.}(2018)\citenamefont
  {Plotkin-Swing}, \citenamefont {Gochnauer}, \citenamefont {McAlpine},
  \citenamefont {Cooper}, \citenamefont {Jamison},\ and\ \citenamefont
  {Gupta}}]{plotkin-swing18}%
  \BibitemOpen
  \bibfield  {author} {\bibinfo {author} {\bibfnamefont {B.}~\bibnamefont
  {Plotkin-Swing}}, \bibinfo {author} {\bibfnamefont {D.}~\bibnamefont
  {Gochnauer}}, \bibinfo {author} {\bibfnamefont {K.~E.}\ \bibnamefont
  {McAlpine}}, \bibinfo {author} {\bibfnamefont {E.~S.}\ \bibnamefont
  {Cooper}}, \bibinfo {author} {\bibfnamefont {A.~O.}\ \bibnamefont
  {Jamison}},\ and\ \bibinfo {author} {\bibfnamefont {S.}~\bibnamefont
  {Gupta}},\ }\bibfield  {title} {\bibinfo {title} {Three-path atom
  interferometry with large momentum separation},\ }\href
  {https://doi.org/10.1103/PhysRevLett.121.133201} {\bibfield  {journal}
  {\bibinfo  {journal} {Phys. Rev. Lett.}\ }\textbf {\bibinfo {volume} {121}},\
  \bibinfo {pages} {133201} (\bibinfo {year} {2018})}\BibitemShut {NoStop}%
\bibitem [{\citenamefont {Mazzoni}\ \emph {et~al.}(2015)\citenamefont
  {Mazzoni}, \citenamefont {Zhang}, \citenamefont {Del~Aguila}, \citenamefont
  {Salvi}, \citenamefont {Poli},\ and\ \citenamefont {Tino}}]{mazzoni15}%
  \BibitemOpen
  \bibfield  {author} {\bibinfo {author} {\bibfnamefont {T.}~\bibnamefont
  {Mazzoni}}, \bibinfo {author} {\bibfnamefont {X.}~\bibnamefont {Zhang}},
  \bibinfo {author} {\bibfnamefont {R.}~\bibnamefont {Del~Aguila}}, \bibinfo
  {author} {\bibfnamefont {L.}~\bibnamefont {Salvi}}, \bibinfo {author}
  {\bibfnamefont {N.}~\bibnamefont {Poli}},\ and\ \bibinfo {author}
  {\bibfnamefont {G.~M.}\ \bibnamefont {Tino}},\ }\bibfield  {title} {\bibinfo
  {title} {Large-momentum-transfer bragg interferometer with strontium atoms},\
  }\href {https://doi.org/10.1103/PhysRevA.92.053619} {\bibfield  {journal}
  {\bibinfo  {journal} {Phys. Rev. A}\ }\textbf {\bibinfo {volume} {92}},\
  \bibinfo {pages} {053619} (\bibinfo {year} {2015})}\BibitemShut {NoStop}%
\bibitem [{\citenamefont {Roy}\ \emph {et~al.}(2016)\citenamefont {Roy},
  \citenamefont {Green}, \citenamefont {Bowler},\ and\ \citenamefont
  {Gupta}}]{roy16}%
  \BibitemOpen
  \bibfield  {author} {\bibinfo {author} {\bibfnamefont {R.}~\bibnamefont
  {Roy}}, \bibinfo {author} {\bibfnamefont {A.}~\bibnamefont {Green}}, \bibinfo
  {author} {\bibfnamefont {R.}~\bibnamefont {Bowler}},\ and\ \bibinfo {author}
  {\bibfnamefont {S.}~\bibnamefont {Gupta}},\ }\bibfield  {title} {\bibinfo
  {title} {Rapid cooling to quantum degeneracy in dynamically shaped atom
  traps},\ }\href {https://doi.org/10.1103/PhysRevA.93.043403} {\bibfield
  {journal} {\bibinfo  {journal} {Phys. Rev. A}\ }\textbf {\bibinfo {volume}
  {93}},\ \bibinfo {pages} {043403} (\bibinfo {year} {2016})}\BibitemShut
  {NoStop}%
\bibitem [{\citenamefont {Alden}\ \emph {et~al.}(2014)\citenamefont {Alden},
  \citenamefont {Moore},\ and\ \citenamefont {Leanhardt}}]{alden14}%
  \BibitemOpen
  \bibfield  {author} {\bibinfo {author} {\bibfnamefont {E.~A.}\ \bibnamefont
  {Alden}}, \bibinfo {author} {\bibfnamefont {K.~R.}\ \bibnamefont {Moore}},\
  and\ \bibinfo {author} {\bibfnamefont {A.~E.}\ \bibnamefont {Leanhardt}},\
  }\bibfield  {title} {\bibinfo {title} {Two-photon {E1--M1} optical clock},\
  }\href@noop {} {\bibfield  {journal} {\bibinfo  {journal} {Phys. Rev. A}\
  }\textbf {\bibinfo {volume} {90}},\ \bibinfo {pages} {012523} (\bibinfo
  {year} {2014})}\BibitemShut {NoStop}%
\bibitem [{\citenamefont {Damour}(2012)}]{damour12}%
  \BibitemOpen
  \bibfield  {author} {\bibinfo {author} {\bibfnamefont {T.}~\bibnamefont
  {Damour}},\ }\bibfield  {title} {\bibinfo {title} {Theoretical aspects of the
  equivalence principle},\ }\href@noop {} {\bibfield  {journal} {\bibinfo
  {journal} {Class. Quant. Grav.}\ }\textbf {\bibinfo {volume} {29}},\ \bibinfo
  {pages} {184001} (\bibinfo {year} {2012})}\BibitemShut {NoStop}%
\bibitem [{\citenamefont {Damour}\ and\ \citenamefont
  {Donoghue}(2010)}]{damour10a}%
  \BibitemOpen
  \bibfield  {author} {\bibinfo {author} {\bibfnamefont {T.}~\bibnamefont
  {Damour}}\ and\ \bibinfo {author} {\bibfnamefont {J.~F.}\ \bibnamefont
  {Donoghue}},\ }\bibfield  {title} {\bibinfo {title} {Equivalence principle
  violations and couplings of a light dilaton},\ }\href
  {https://doi.org/10.1103/PhysRevD.82.084033} {\bibfield  {journal} {\bibinfo
  {journal} {Phys. Rev. D}\ }\textbf {\bibinfo {volume} {82}},\ \bibinfo
  {pages} {084033} (\bibinfo {year} {2010})}\BibitemShut {NoStop}%
\bibitem [{Note1()}]{Note1}%
  \BibitemOpen
  \bibinfo {note} {This relation has previously been found for the comparison
  of independent clocks by employing energy conservation arguments (and Lorentz
  invariance) \cite {nordtvedt75,wolf16}.}\BibitemShut {Stop}%
\bibitem [{Note2()}]{Note2}%
  \BibitemOpen
  \bibinfo {note} {For simplicity we have assumed that $\Delta \protect \mathbf
  {g}$ was time independent during each shot. Otherwise its contribution to
  Eq.~\protect \textup {\hbox {\mathsurround \z@ \protect \normalfont
  (\ignorespaces \ref {eq:laser_phase_shift}\unskip \@@italiccorr )}} would be
  replaced by a linear combination of double time integrals, but all the
  relevant conclusions would remain unchanged.}\BibitemShut {Stop}%
\bibitem [{\citenamefont {Peters}\ \emph {et~al.}(2001)\citenamefont {Peters},
  \citenamefont {Chung},\ and\ \citenamefont {Chu}}]{peters01}%
  \BibitemOpen
  \bibfield  {author} {\bibinfo {author} {\bibfnamefont {A.}~\bibnamefont
  {Peters}}, \bibinfo {author} {\bibfnamefont {K.~Y.}\ \bibnamefont {Chung}},\
  and\ \bibinfo {author} {\bibfnamefont {S.}~\bibnamefont {Chu}},\ }\bibfield
  {title} {\bibinfo {title} {High-precision gravity measurements using atom
  interferometry},\ }\href {http://stacks.iop.org/0026-1394/38/i=1/a=4}
  {\bibfield  {journal} {\bibinfo  {journal} {Metrologia}\ }\textbf {\bibinfo
  {volume} {38}},\ \bibinfo {pages} {25} (\bibinfo {year} {2001})}\BibitemShut
  {NoStop}%
\bibitem [{Note3()}]{Note3}%
  \BibitemOpen
  \bibinfo {note} {Note that any terms depending on the initial position and
  velocity of the mirror cancel out and one is just left with those depending
  on its acceleration.}\BibitemShut {Stop}%
\bibitem [{\citenamefont {Kovachy}\ \emph
  {et~al.}(2015{\natexlab{b}})\citenamefont {Kovachy}, \citenamefont {Hogan},
  \citenamefont {Sugarbaker}, \citenamefont {Dickerson}, \citenamefont
  {Donnelly}, \citenamefont {Overstreet},\ and\ \citenamefont
  {Kasevich}}]{kovachy15a}%
  \BibitemOpen
  \bibfield  {author} {\bibinfo {author} {\bibfnamefont {T.}~\bibnamefont
  {Kovachy}}, \bibinfo {author} {\bibfnamefont {J.~M.}\ \bibnamefont {Hogan}},
  \bibinfo {author} {\bibfnamefont {A.}~\bibnamefont {Sugarbaker}}, \bibinfo
  {author} {\bibfnamefont {S.~M.}\ \bibnamefont {Dickerson}}, \bibinfo {author}
  {\bibfnamefont {C.~A.}\ \bibnamefont {Donnelly}}, \bibinfo {author}
  {\bibfnamefont {C.}~\bibnamefont {Overstreet}},\ and\ \bibinfo {author}
  {\bibfnamefont {M.~A.}\ \bibnamefont {Kasevich}},\ }\bibfield  {title}
  {\bibinfo {title} {Matter wave lensing to picokelvin temperatures},\ }\href
  {https://doi.org/10.1103/PhysRevLett.114.143004} {\bibfield  {journal}
  {\bibinfo  {journal} {Phys. Rev. Lett.}\ }\textbf {\bibinfo {volume} {114}},\
  \bibinfo {pages} {143004} (\bibinfo {year} {2015}{\natexlab{b}})}\BibitemShut
  {NoStop}%
\bibitem [{Note4()}]{Note4}%
  \BibitemOpen
  \bibinfo {note} {If necessary, one could slightly shift all the pulse times
  of the Rb interferometer by the same small amount because high-frequency
  vibrations are typically much better suppressed by the mirror's vibration
  isolation system.}\BibitemShut {Stop}%
\bibitem [{\citenamefont {Takasu}\ \emph {et~al.}(2003)\citenamefont {Takasu},
  \citenamefont {Maki}, \citenamefont {Komori}, \citenamefont {Takano},
  \citenamefont {Honda}, \citenamefont {Kumakura}, \citenamefont {Yabuzaki},\
  and\ \citenamefont {Takahashi}}]{takasu03}%
  \BibitemOpen
  \bibfield  {author} {\bibinfo {author} {\bibfnamefont {Y.}~\bibnamefont
  {Takasu}}, \bibinfo {author} {\bibfnamefont {K.}~\bibnamefont {Maki}},
  \bibinfo {author} {\bibfnamefont {K.}~\bibnamefont {Komori}}, \bibinfo
  {author} {\bibfnamefont {T.}~\bibnamefont {Takano}}, \bibinfo {author}
  {\bibfnamefont {K.}~\bibnamefont {Honda}}, \bibinfo {author} {\bibfnamefont
  {M.}~\bibnamefont {Kumakura}}, \bibinfo {author} {\bibfnamefont
  {T.}~\bibnamefont {Yabuzaki}},\ and\ \bibinfo {author} {\bibfnamefont
  {Y.}~\bibnamefont {Takahashi}},\ }\bibfield  {title} {\bibinfo {title}
  {Spin-singlet {Bose-Einstein} condensation of two-electron atoms},\ }\href
  {https://doi.org/10.1103/PhysRevLett.91.040404} {\bibfield  {journal}
  {\bibinfo  {journal} {Phys. Rev. Lett.}\ }\textbf {\bibinfo {volume} {91}},\
  \bibinfo {pages} {040404} (\bibinfo {year} {2003})}\BibitemShut {NoStop}%
\bibitem [{\citenamefont {Fukuhara}\ \emph {et~al.}(2007)\citenamefont
  {Fukuhara}, \citenamefont {Sugawa},\ and\ \citenamefont
  {Takahashi}}]{fukuhara07}%
  \BibitemOpen
  \bibfield  {author} {\bibinfo {author} {\bibfnamefont {T.}~\bibnamefont
  {Fukuhara}}, \bibinfo {author} {\bibfnamefont {S.}~\bibnamefont {Sugawa}},\
  and\ \bibinfo {author} {\bibfnamefont {Y.}~\bibnamefont {Takahashi}},\
  }\bibfield  {title} {\bibinfo {title} {{Bose-Einstein} condensation of an
  ytterbium isotope},\ }\href {https://doi.org/10.1103/PhysRevA.76.051604}
  {\bibfield  {journal} {\bibinfo  {journal} {Phys. Rev. A}\ }\textbf {\bibinfo
  {volume} {76}},\ \bibinfo {pages} {051604} (\bibinfo {year}
  {2007})}\BibitemShut {NoStop}%
\bibitem [{\citenamefont {D\"orscher}\ \emph {et~al.}(2013)\citenamefont
  {D\"orscher}, \citenamefont {Thobe}, \citenamefont {Hundt}, \citenamefont
  {Kochanke}, \citenamefont {Le~Targat}, \citenamefont {Windpassinger},
  \citenamefont {Becker},\ and\ \citenamefont {Sengstock}}]{doerscher13}%
  \BibitemOpen
  \bibfield  {author} {\bibinfo {author} {\bibfnamefont {S.}~\bibnamefont
  {D\"orscher}}, \bibinfo {author} {\bibfnamefont {A.}~\bibnamefont {Thobe}},
  \bibinfo {author} {\bibfnamefont {B.}~\bibnamefont {Hundt}}, \bibinfo
  {author} {\bibfnamefont {A.}~\bibnamefont {Kochanke}}, \bibinfo {author}
  {\bibfnamefont {R.}~\bibnamefont {Le~Targat}}, \bibinfo {author}
  {\bibfnamefont {P.}~\bibnamefont {Windpassinger}}, \bibinfo {author}
  {\bibfnamefont {C.}~\bibnamefont {Becker}},\ and\ \bibinfo {author}
  {\bibfnamefont {K.}~\bibnamefont {Sengstock}},\ }\bibfield  {title} {\bibinfo
  {title} {Creation of quantum-degenerate gases of ytterbium in a compact
  2d-/3d-magneto-optical trap setup},\ }\href
  {https://doi.org/10.1063/1.4802682} {\bibfield  {journal} {\bibinfo
  {journal} {Rev. Sci. Instrum.}\ }\textbf {\bibinfo {volume} {84}},\ \bibinfo
  {pages} {043109} (\bibinfo {year} {2013})}\BibitemShut {NoStop}%
\bibitem [{\citenamefont {M\"{u}ntinga}\ \emph {et~al.}(2013)\citenamefont
  {M\"{u}ntinga}, \citenamefont {Ahlers}, \citenamefont {Krutzik},
  \citenamefont {Wenzlawski}, \citenamefont {Arnold}, \citenamefont {Becker},
  \citenamefont {Bongs}, \citenamefont {Dittus}, \citenamefont {Duncker},
  \citenamefont {Gaaloul}, \citenamefont {Gherasim}, \citenamefont {Giese},
  \citenamefont {Grzeschik}, \citenamefont {H\"{a}nsch}, \citenamefont
  {Hellmig}, \citenamefont {Herr}, \citenamefont {Herrmann}, \citenamefont
  {Kajari}, \citenamefont {Kleinert}, \citenamefont {L\"{a}mmerzahl},
  \citenamefont {Lewoczko-Adamczyk}, \citenamefont {Malcolm}, \citenamefont
  {Meyer}, \citenamefont {Nolte}, \citenamefont {Peters}, \citenamefont {Popp},
  \citenamefont {Reichel}, \citenamefont {Roura}, \citenamefont {Rudolph},
  \citenamefont {Schiemangk}, \citenamefont {Schneider}, \citenamefont
  {Seidel}, \citenamefont {Sengstock}, \citenamefont {Tamma}, \citenamefont
  {Valenzuela}, \citenamefont {Vogel}, \citenamefont {Walser}, \citenamefont
  {Wendrich}, \citenamefont {Windpassinger}, \citenamefont {Zeller},
  \citenamefont {van Zoest}, \citenamefont {Ertmer}, \citenamefont {Schleich},\
  and\ \citenamefont {Rasel}}]{muentinga13}%
  \BibitemOpen
  \bibfield  {author} {\bibinfo {author} {\bibfnamefont {H.}~\bibnamefont
  {M\"{u}ntinga}}, \bibinfo {author} {\bibfnamefont {H.}~\bibnamefont
  {Ahlers}}, \bibinfo {author} {\bibfnamefont {M.}~\bibnamefont {Krutzik}},
  \bibinfo {author} {\bibfnamefont {A.}~\bibnamefont {Wenzlawski}}, \bibinfo
  {author} {\bibfnamefont {S.}~\bibnamefont {Arnold}}, \bibinfo {author}
  {\bibfnamefont {D.}~\bibnamefont {Becker}}, \bibinfo {author} {\bibfnamefont
  {K.}~\bibnamefont {Bongs}}, \bibinfo {author} {\bibfnamefont
  {H.}~\bibnamefont {Dittus}}, \bibinfo {author} {\bibfnamefont
  {H.}~\bibnamefont {Duncker}}, \bibinfo {author} {\bibfnamefont
  {N.}~\bibnamefont {Gaaloul}}, \bibinfo {author} {\bibfnamefont
  {C.}~\bibnamefont {Gherasim}}, \bibinfo {author} {\bibfnamefont
  {E.}~\bibnamefont {Giese}}, \bibinfo {author} {\bibfnamefont
  {C.}~\bibnamefont {Grzeschik}}, \bibinfo {author} {\bibfnamefont {T.~W.}\
  \bibnamefont {H\"{a}nsch}}, \bibinfo {author} {\bibfnamefont
  {O.}~\bibnamefont {Hellmig}}, \bibinfo {author} {\bibfnamefont
  {W.}~\bibnamefont {Herr}}, \bibinfo {author} {\bibfnamefont {S.}~\bibnamefont
  {Herrmann}}, \bibinfo {author} {\bibfnamefont {E.}~\bibnamefont {Kajari}},
  \bibinfo {author} {\bibfnamefont {S.}~\bibnamefont {Kleinert}}, \bibinfo
  {author} {\bibfnamefont {C.}~\bibnamefont {L\"{a}mmerzahl}}, \bibinfo
  {author} {\bibfnamefont {W.}~\bibnamefont {Lewoczko-Adamczyk}}, \bibinfo
  {author} {\bibfnamefont {J.}~\bibnamefont {Malcolm}}, \bibinfo {author}
  {\bibfnamefont {N.}~\bibnamefont {Meyer}}, \bibinfo {author} {\bibfnamefont
  {R.}~\bibnamefont {Nolte}}, \bibinfo {author} {\bibfnamefont
  {A.}~\bibnamefont {Peters}}, \bibinfo {author} {\bibfnamefont
  {M.}~\bibnamefont {Popp}}, \bibinfo {author} {\bibfnamefont {J.}~\bibnamefont
  {Reichel}}, \bibinfo {author} {\bibfnamefont {A.}~\bibnamefont {Roura}},
  \bibinfo {author} {\bibfnamefont {J.}~\bibnamefont {Rudolph}}, \bibinfo
  {author} {\bibfnamefont {M.}~\bibnamefont {Schiemangk}}, \bibinfo {author}
  {\bibfnamefont {M.}~\bibnamefont {Schneider}}, \bibinfo {author}
  {\bibfnamefont {S.~T.}\ \bibnamefont {Seidel}}, \bibinfo {author}
  {\bibfnamefont {K.}~\bibnamefont {Sengstock}}, \bibinfo {author}
  {\bibfnamefont {V.}~\bibnamefont {Tamma}}, \bibinfo {author} {\bibfnamefont
  {T.}~\bibnamefont {Valenzuela}}, \bibinfo {author} {\bibfnamefont
  {A.}~\bibnamefont {Vogel}}, \bibinfo {author} {\bibfnamefont
  {R.}~\bibnamefont {Walser}}, \bibinfo {author} {\bibfnamefont
  {T.}~\bibnamefont {Wendrich}}, \bibinfo {author} {\bibfnamefont
  {P.}~\bibnamefont {Windpassinger}}, \bibinfo {author} {\bibfnamefont
  {W.}~\bibnamefont {Zeller}}, \bibinfo {author} {\bibfnamefont
  {T.}~\bibnamefont {van Zoest}}, \bibinfo {author} {\bibfnamefont
  {W.}~\bibnamefont {Ertmer}}, \bibinfo {author} {\bibfnamefont {W.~P.}\
  \bibnamefont {Schleich}},\ and\ \bibinfo {author} {\bibfnamefont {E.~M.}\
  \bibnamefont {Rasel}},\ }\bibfield  {title} {\bibinfo {title} {Interferometry
  with {Bose-Einstein} condensates in microgravity},\ }\href
  {https://doi.org/10.1103/PhysRevLett.110.093602} {\bibfield  {journal}
  {\bibinfo  {journal} {Phys. Rev. Lett.}\ }\textbf {\bibinfo {volume} {110}},\
  \bibinfo {pages} {093602} (\bibinfo {year} {2013})}\BibitemShut {NoStop}%
\bibitem [{\citenamefont {Chiow}\ \emph {et~al.}(2011)\citenamefont {Chiow},
  \citenamefont {Kovachy}, \citenamefont {Chien},\ and\ \citenamefont
  {Kasevich}}]{chiow11}%
  \BibitemOpen
  \bibfield  {author} {\bibinfo {author} {\bibfnamefont {S.-w.}\ \bibnamefont
  {Chiow}}, \bibinfo {author} {\bibfnamefont {T.}~\bibnamefont {Kovachy}},
  \bibinfo {author} {\bibfnamefont {H.-C.}\ \bibnamefont {Chien}},\ and\
  \bibinfo {author} {\bibfnamefont {M.~A.}\ \bibnamefont {Kasevich}},\
  }\bibfield  {title} {\bibinfo {title} {$102\ensuremath{\hbar}k$ large area
  atom interferometers},\ }\href
  {https://doi.org/10.1103/PhysRevLett.107.130403} {\bibfield  {journal}
  {\bibinfo  {journal} {Phys. Rev. Lett.}\ }\textbf {\bibinfo {volume} {107}},\
  \bibinfo {pages} {130403} (\bibinfo {year} {2011})}\BibitemShut {NoStop}%
\bibitem [{\citenamefont {Gebbe}\ \emph {et~al.}()\citenamefont {Gebbe} \emph
  {et~al.}}]{gebbe19}%
  \BibitemOpen
  \bibfield  {author} {\bibinfo {author} {\bibfnamefont {M.}~\bibnamefont
  {Gebbe}} \emph {et~al.},\ }\bibfield  {title} {\bibinfo {title} {Twin-lattice
  atom interferometry},\ }\href {https://arxiv.org/abs/1907.08416} {\bibinfo
  {journal} {\texttt{arXiv:1907.08416}}\ }\BibitemShut {NoStop}%
\bibitem [{\citenamefont {Lan}\ \emph {et~al.}(2012)\citenamefont {Lan},
  \citenamefont {Kuan}, \citenamefont {Estey}, \citenamefont {Haslinger},\ and\
  \citenamefont {M\"{u}ller}}]{lan12}%
  \BibitemOpen
\bibfield  {journal} {  }\bibfield  {author} {\bibinfo {author} {\bibfnamefont
  {S.-Y.}\ \bibnamefont {Lan}}, \bibinfo {author} {\bibfnamefont {P.-C.}\
  \bibnamefont {Kuan}}, \bibinfo {author} {\bibfnamefont {B.}~\bibnamefont
  {Estey}}, \bibinfo {author} {\bibfnamefont {P.}~\bibnamefont {Haslinger}},\
  and\ \bibinfo {author} {\bibfnamefont {H.}~\bibnamefont {M\"{u}ller}},\
  }\bibfield  {title} {\bibinfo {title} {Influence of the {C}oriolis force in
  atom interferometry},\ }\href
  {https://doi.org/10.1103/PhysRevLett.108.090402} {\bibfield  {journal}
  {\bibinfo  {journal} {Phys. Rev. Lett.}\ }\textbf {\bibinfo {volume} {108}},\
  \bibinfo {pages} {090402} (\bibinfo {year} {2012})}\BibitemShut {NoStop}%
\bibitem [{\citenamefont {Dickerson}\ \emph {et~al.}(2013)\citenamefont
  {Dickerson}, \citenamefont {Hogan}, \citenamefont {Sugarbaker}, \citenamefont
  {Johnson},\ and\ \citenamefont {Kasevich}}]{dickerson13}%
  \BibitemOpen
  \bibfield  {author} {\bibinfo {author} {\bibfnamefont {S.~M.}\ \bibnamefont
  {Dickerson}}, \bibinfo {author} {\bibfnamefont {J.~M.}\ \bibnamefont
  {Hogan}}, \bibinfo {author} {\bibfnamefont {A.}~\bibnamefont {Sugarbaker}},
  \bibinfo {author} {\bibfnamefont {D.~M.~S.}\ \bibnamefont {Johnson}},\ and\
  \bibinfo {author} {\bibfnamefont {M.~A.}\ \bibnamefont {Kasevich}},\
  }\bibfield  {title} {\bibinfo {title} {Multiaxis inertial sensing with
  long-time point source atom interferometry},\ }\href
  {https://doi.org/10.1103/PhysRevLett.111.083001} {\bibfield  {journal}
  {\bibinfo  {journal} {Phys. Rev. Lett.}\ }\textbf {\bibinfo {volume} {111}},\
  \bibinfo {pages} {083001} (\bibinfo {year} {2013})}\BibitemShut {NoStop}%
\bibitem [{\citenamefont {Roura}(2017)}]{roura17a}%
  \BibitemOpen
  \bibfield  {author} {\bibinfo {author} {\bibfnamefont {A.}~\bibnamefont
  {Roura}},\ }\bibfield  {title} {\bibinfo {title} {Circumventing
  {Heisenberg's} uncertainty principle in atom interferometry tests of the
  equivalence principle},\ }\href
  {https://doi.org/10.1103/PhysRevLett.118.160401} {\bibfield  {journal}
  {\bibinfo  {journal} {Phys. Rev. Lett.}\ }\textbf {\bibinfo {volume} {118}},\
  \bibinfo {pages} {160401} (\bibinfo {year} {2017})}\BibitemShut {NoStop}%
\bibitem [{\citenamefont {D'Amico}\ \emph {et~al.}(2017)\citenamefont
  {D'Amico}, \citenamefont {Rosi}, \citenamefont {Zhan}, \citenamefont
  {Cacciapuoti}, \citenamefont {Fattori},\ and\ \citenamefont
  {Tino}}]{d_amico17}%
  \BibitemOpen
  \bibfield  {author} {\bibinfo {author} {\bibfnamefont {G.}~\bibnamefont
  {D'Amico}}, \bibinfo {author} {\bibfnamefont {G.}~\bibnamefont {Rosi}},
  \bibinfo {author} {\bibfnamefont {S.}~\bibnamefont {Zhan}}, \bibinfo {author}
  {\bibfnamefont {L.}~\bibnamefont {Cacciapuoti}}, \bibinfo {author}
  {\bibfnamefont {M.}~\bibnamefont {Fattori}},\ and\ \bibinfo {author}
  {\bibfnamefont {G.~M.}\ \bibnamefont {Tino}},\ }\bibfield  {title} {\bibinfo
  {title} {Canceling the gravity gradient phase shift in atom interferometry},\
  }\href@noop {} {\bibfield  {journal} {\bibinfo  {journal} {Phys. Rev. Lett.}\
  }\textbf {\bibinfo {volume} {119}},\ \bibinfo {pages} {253201} (\bibinfo
  {year} {2017})}\BibitemShut {NoStop}%
\bibitem [{\citenamefont {Overstreet}\ \emph {et~al.}(2018)\citenamefont
  {Overstreet}, \citenamefont {Asenbaum}, \citenamefont {Kovachy},
  \citenamefont {Notermans}, \citenamefont {Hogan},\ and\ \citenamefont
  {Kasevich}}]{overstreet18}%
  \BibitemOpen
  \bibfield  {author} {\bibinfo {author} {\bibfnamefont {C.}~\bibnamefont
  {Overstreet}}, \bibinfo {author} {\bibfnamefont {P.}~\bibnamefont
  {Asenbaum}}, \bibinfo {author} {\bibfnamefont {T.}~\bibnamefont {Kovachy}},
  \bibinfo {author} {\bibfnamefont {R.}~\bibnamefont {Notermans}}, \bibinfo
  {author} {\bibfnamefont {J.~M.}\ \bibnamefont {Hogan}},\ and\ \bibinfo
  {author} {\bibfnamefont {M.~A.}\ \bibnamefont {Kasevich}},\ }\bibfield
  {title} {\bibinfo {title} {Effective inertial frame in an atom
  interferometric test of the equivalence principle},\ }\href
  {https://doi.org/10.1103/PhysRevLett.120.183604} {\bibfield  {journal}
  {\bibinfo  {journal} {Phys. Rev. Lett.}\ }\textbf {\bibinfo {volume} {120}},\
  \bibinfo {pages} {183604} (\bibinfo {year} {2018})}\BibitemShut {NoStop}%
\bibitem [{\citenamefont {Geiger}\ and\ \citenamefont
  {Trupke}(2018)}]{geiger18}%
  \BibitemOpen
  \bibfield  {author} {\bibinfo {author} {\bibfnamefont {R.}~\bibnamefont
  {Geiger}}\ and\ \bibinfo {author} {\bibfnamefont {M.}~\bibnamefont
  {Trupke}},\ }\bibfield  {title} {\bibinfo {title} {Proposal for a quantum
  test of the weak equivalence principle with entangled atomic species},\
  }\href {https://doi.org/10.1103/PhysRevLett.120.043602} {\bibfield  {journal}
  {\bibinfo  {journal} {Phys. Rev. Lett.}\ }\textbf {\bibinfo {volume} {120}},\
  \bibinfo {pages} {043602} (\bibinfo {year} {2018})}\BibitemShut {NoStop}%
\bibitem [{\citenamefont {{VLBAI at HITec}}()}]{hitec}%
  \BibitemOpen
  \bibfield  {author} {\bibinfo {author} {\bibnamefont {{VLBAI at HITec}}},\
  }\href
  {https://www.hitec.uni-hannover.de/en/large-scale-equipment/atomic-fountain}
  {\bibinfo  {journal}
  {https://www.hitec.uni-hannover.de/\linebreak[3]en/large-scale-equipment/atomic-fountain}\
  }\BibitemShut {NoStop}%
\bibitem [{\citenamefont {L\'ev\`eque}\ \emph {et~al.}(2009)\citenamefont
  {L\'ev\`eque}, \citenamefont {Gauguet}, \citenamefont {Michaud},
  \citenamefont {Pereira Dos~Santos},\ and\ \citenamefont
  {Landragin}}]{leveque09}%
  \BibitemOpen
\bibfield  {journal} {  }\bibfield  {author} {\bibinfo {author} {\bibfnamefont
  {T.}~\bibnamefont {L\'ev\`eque}}, \bibinfo {author} {\bibfnamefont
  {A.}~\bibnamefont {Gauguet}}, \bibinfo {author} {\bibfnamefont
  {F.}~\bibnamefont {Michaud}}, \bibinfo {author} {\bibfnamefont
  {F.}~\bibnamefont {Pereira Dos~Santos}},\ and\ \bibinfo {author}
  {\bibfnamefont {A.}~\bibnamefont {Landragin}},\ }\bibfield  {title} {\bibinfo
  {title} {Enhancing the area of a {R}aman atom interferometer using a
  versatile double-diffraction technique},\ }\href
  {https://doi.org/10.1103/PhysRevLett.103.080405} {\bibfield  {journal}
  {\bibinfo  {journal} {Phys. Rev. Lett.}\ }\textbf {\bibinfo {volume} {103}},\
  \bibinfo {pages} {080405} (\bibinfo {year} {2009})}\BibitemShut {NoStop}%
\bibitem [{Note5()}]{Note5}%
  \BibitemOpen
  \bibinfo {note} {For open interferometers there is an additional contribution
  corresponding to the so-called separation phase $\delta \phi _\protect \text
  {sep}$ \cite {roura20a}. However, in this paper we focus on closed
  interferomters, i.e., with no relative displacement between the interfering
  wave packets.}\BibitemShut {Stop}%
\bibitem [{\citenamefont {Nordtvedt}(1975)}]{nordtvedt75}%
  \BibitemOpen
  \bibfield  {author} {\bibinfo {author} {\bibfnamefont {K.}~\bibnamefont
  {Nordtvedt}},\ }\bibfield  {title} {\bibinfo {title} {Quantitative
  relationship between clock gravitational ``red-shift'' violations and
  nonuniversality of free-fall rates in nonmetric theories of gravity},\ }\href
  {https://doi.org/10.1103/PhysRevD.11.245} {\bibfield  {journal} {\bibinfo
  {journal} {Phys. Rev. D}\ }\textbf {\bibinfo {volume} {11}},\ \bibinfo
  {pages} {245} (\bibinfo {year} {1975})}\BibitemShut {NoStop}%
\bibitem [{\citenamefont {Wolf}\ and\ \citenamefont {Blanchet}(2016)}]{wolf16}%
  \BibitemOpen
  \bibfield  {author} {\bibinfo {author} {\bibfnamefont {P.}~\bibnamefont
  {Wolf}}\ and\ \bibinfo {author} {\bibfnamefont {L.}~\bibnamefont
  {Blanchet}},\ }\bibfield  {title} {\bibinfo {title} {Analysis of sun/moon
  gravitational redshift tests with the {STE-QUEST} space mission},\
  }\href@noop {} {\bibfield  {journal} {\bibinfo  {journal} {Class. Quant.
  Grav.}\ }\textbf {\bibinfo {volume} {33}},\ \bibinfo {pages} {035012}
  (\bibinfo {year} {2016})}\BibitemShut {NoStop}%
\end{thebibliography}%

\end{document}